\documentclass[superscriptaddress,dvipsnames,nofootinbib,amsmath,amssymb,aps,prd,floatfix,reprint]{revtex4-2}

\usepackage[titletoc,toc,title]{appendix}
\usepackage{mathtools}
\usepackage{amsmath, amsfonts,bm}
\usepackage[utf8]{inputenc}
\usepackage{xcolor}
\usepackage{graphicx}
\usepackage{float}
\usepackage{subfig}
\usepackage{comment}
\newcommand\myshade{85}
\colorlet{mylinkcolor}{RoyalPurple}
\colorlet{mycitecolor}{WildStrawberry}
\colorlet{myurlcolor}{BlueViolet}
\usepackage{orcidlink}

\usepackage[normalem]{ulem}

\usepackage{hyperref}
\usepackage[capitalise]{cleveref}

\hypersetup{
  linkcolor  = mylinkcolor!\myshade!black,
  citecolor  = mycitecolor!\myshade!black,
  urlcolor   = myurlcolor!\myshade!black,
  colorlinks = true,
}

\DeclareMathAlphabet{\mathup}{OT1}{\familydefault}{m}{n}

\newcommand{\be}{\begin{equation}} 
\newcommand{\ee}{\end{equation}}

\usepackage{array}
\newcommand{\PreserveBackslash}[1]{\let\temp=\\#1\let\\=\temp}
\newcolumntype{C}[1]{>{\PreserveBackslash\centering}p{#1}}
\newcolumntype{R}[1]{>{\PreserveBackslash\raggedleft}p{#1}}
\newcolumntype{L}[1]{>{\PreserveBackslash\raggedright}p{#1}}

\usepackage{multirow}
\usepackage{bm}


\crefname{equation}{Eq.}{Eqs.}
\crefname{section}{Section}{Sections}
\crefname{figure}{Fig.}{Figs.}
\crefname{table}{Table}{Tables}
\crefname{appendix}{Appendix}{Appendices}
\Crefname{figure}{Figure}{Figures}
\Crefname{equation}{Equation}{Equations}
\Crefname{section}{Section}{Sections}
\Crefname{table}{Table}{Tables}

\definecolor{llgray}{gray}{0.93}
\definecolor{lgray}{gray}{0.83}
\definecolor{deepmagenta}{rgb}{0.8, 0.0, 0.8}
\definecolor{ballblue}{rgb}{0.13, 0.67, 0.8}
\definecolor{celestialblue}{rgb}{0.29, 0.59, 0.82}
\definecolor{RedWine}{rgb}{0.743,0,0}
\definecolor{DarkGreen}{rgb}{0,0.6,0}

\usepackage{fontawesome5}
\newcommand{\github}[1]{\href{#1}{\faGithub}
}

\begin{document}

\title{New multiprobe analysis of modified gravity and evolving dark energy}

\author{Zhiyu Lu\orcidlink{0009-0001-3701-6650}}
\email{zhiyulu@mail.ustc.edu.cn}
\affiliation{Department of Astronomy, School of Physical Sciences, University of Science and Technology of China, Hefei, Anhui 230026, China}
\affiliation{CAS Key Laboratory for Research in Galaxies
and Cosmology, School of Astronomy and Space Science, University of Science and Technology of China, Hefei, Anhui 230026, China}

\author{Th\'eo Simon\orcidlink{0000-0001-7858-6441}}
 \email{theo.simon@umontpellier.fr}
 
\affiliation{Laboratoire Univers et Particules de Montpellier (LUPM), Centre national de la recherche scientifique (CNRS) et Universit\'e de Montpellier, Place Eug\`ene Bataillon, 34095 Montpellier C\'edex 05, France}


\begin{abstract}

We study the $(w_0, \, w_a)$ parametrization of the dark energy (DE) equation of state, with and without the effective field theory of dark energy (EFTofDE) framework to describe the DE perturbations, parametrized here by the braiding parameter $\alpha_B$ and the running of the Planck mass $\alpha_M$.
We combine the EFTofLSS full-shape analysis of the power
spectrum and bispectrum of BOSS data with the tomographic angular power spectra $C_\ell^{gg}$, $C_\ell^{\kappa g}$, $C_\ell^{Tg}$ and $C_\ell^{T\kappa}$, where $g$, $\kappa$ and $T$ stand for the DESI luminous red galaxy map, \textit{Planck} PR4 lensing map and \textit{Planck} PR4 temperature map, respectively.
To analyze these angular power spectra, we go beyond the Limber approximation, allowing us to include large-scale data in $C_\ell^{gg}$.
The combination of all these probes with \textit{Planck} PR4, DESI DR2 BAO and DES Y5 improves the constraint on the 2D posterior distribution of $\{w_0, \, w_a\}$ by $\sim 50 \%$ and increases the preference for evolving dark energy over $\Lambda$ from $3.8 \sigma$ to $4.6 \sigma$.
When we remove BAO and supernovae data, we obtain a hint for evolving dark energy at $2.3 \sigma$.
Regarding the EFTofDE parameters, we improve the constraints on $\alpha_B$ and $\alpha_M$ by $\sim 40 \%$ and $50 \%$ respectively, finding results compatible with general relativity at $\sim 2 \sigma$.
We show that these constraints do not depend on the choice of the BAO and supernovae likelihoods. 

\end{abstract}

\maketitle

\section{Introduction} \label{sec:intro}

Since the discovery of dark energy (DE) in 1998~\cite{SupernovaSearchTeam:1998fmf}, cosmological data seemed to favor the existence of a cosmological constant $\Lambda$, a key element of the standard paradigm of cosmology, the $\Lambda$CDM model.
Cosmological data from cosmic microwave background (CMB)~\cite{Planck:2018vyg}, large-scale structure (LSS)~\cite{BOSS:2016wmc}, or supernovae~\cite{Pan-STARRS1:2017jku} appeared to favor a dark energy equation of state parameter close to $w = -1$, with no compelling evidence for evolving dark energy.
However, since the beginning of 2024, this picture has been challenged by new measurements from the Dark Energy Spectroscopic Instrument (DESI)~\cite{DESI:2024uvr} and supernovae compilations~\cite{Brout:2022vxf,Rubin:2023jdq,DES:2024jxu}.
The first DESI data release (DR1) suggests a mild preference for evolving dark energy when combined with CMB and supernovae data, deviating from a cosmological constant by $2.5\sigma - 3.9\sigma$~\cite{DESI:2024mwx,DESI:2024hhd} (depending on the supernova data used).
This raises concerns about the consistency of supernova data--for instance, Ref.~\cite{Efstathiou:2024xcq} pointed out that a systematic offset in the magnitude between low and high redshift supernovae could mimic an evolving dark energy signal. 
However, with the second DESI data release, a preference for evolving dark energy at $3.1 \sigma$ was observed without supernovae measurements~\cite{DESI:2025zgx}, along with a $2.3\sigma$ discrepancy between DESI DR2 BAO and \textit{Planck} (assuming the $\Lambda$CDM model) in the 2D posterior distribution of $\{ \Omega_m, \, r_d\cdot h \}$, where $r_d$ is the sound horizon at baryon drag.
This discrepancy is exacerbated to $3.7\sigma$ in the SPT + ACT analysis~\cite{SPT-3G:2025bzu}.
We note that this inconsistency can also be mitigated if we vary the sum of the neutrino mass~\cite{Cozzumbo:2025ewt}, increase the value of the optical depth $\tau$~\cite{Jhaveri:2025neg}, or by some early-time dynamics~\cite{Sharma:2025iux,Poulin:2025nfb,Mirpoorian:2025rfp}.

Interestingly, the reconstructed dark energy equation of state crosses $w = -1$ around $z \sim 0.5$, implying a violation of the null energy condition at high redshift. 
Various models have been proposed to explain this behavior and address this challenge. 
For example, the so-called phantom mirage mechanism--where a coupling between quintessence and dark matter produces an apparent evolution in $w$--can mimic dynamical dark energy~\cite{Amendola:1999er,Das:2005yj}. 
We can also invoke axion-like scalar field models~\cite{Liu:2025bss}, scalar field models with a noncanonical kinetic term~\cite{Wolf:2025acj} or non-minimally coupled scalar field models~\cite{Ye:2024ywg}. 
An alternative explanation involves the quintom scenario~\cite{Cai:2025mas,Cai:2009zp,Feng:2004ad,Xia:2007km,Zhao:2012aw,Guo:2004fq,Zhao:2005vj,Xia:2005ge}, in which two scalar fields (one of quintessence type and the other of phantom type) combine to allow a crossing of the $w = -1$ limit without instability.

On the other hand, the potential preference for evolving dark energy is a good motivation for studying modified gravity models: is the evolving dark energy, in fact, the manifestation of an extension or deviation from general relativity~\cite{Yang:2024kdo,Yang:2025mws,Cai:2015emx,Li:2018ixg}?
In the case of a single scalar degree of freedom, the effective theory of dark energy (EFTofDE)~\cite{Creminelli:2008wc,Creminelli:2006xe,Frusciante:2019xia,Bellini:2014fua,Gubitosi:2012hu,Gleyzes:2014rba,Gleyzes:2013ooa} provides a unified description of gravity beyond general relativity, separating background expansion from linear perturbations. The expansion history can be parametrized by an arbitrary equation of state $w(z)$, while the dynamics of perturbations are described by the general time-dependent functions $\alpha_K$, $\alpha_B$, $\alpha_M$, and $\alpha_T$. 
Though no significant derivation from general relativity has been reported so far, a well-known challenge is the correlation between the braiding parameter $\alpha_B$ and the lensing amplitude $A_L$. The so-called “lensing anomaly” in CMB data can shift $\alpha_B$ away from zero by up to $2\sigma$~\cite{Lu:2025gki,Ishak:2024jhs,Noller:2018wyv,Planck:2015bue}, although the significance of this anomaly is reduced in the \textit{Planck} 2020 analysis compared to \textit{Planck} 2018~\cite{Tristram:2023haj}.

{Modified gravity alters the evolution of metric perturbations at late times through the Poisson equation and the Weyl potential, which can be directly probed by cosmological observations that are sensitive to the late-time perturbed Universe. 
For instance, it will affect the matter overdensity field (and therefore the galaxy overdensity field), the
CMB gravitational lensing, as well as the CMB temperature power spectrum at large angular scales~\cite{Das:2013sca} [through the integrated Sachs-Wolfe (ISW) effect].
In order to constrain the EFTofDE parameters, we consider in this paper cross-correlations between these three observables, namely, $C_\ell^{\kappa g}$, $C_\ell^{T g}$, and $C_\ell^{T \kappa}$, corresponding, respectively, to the cross-angular power spectra between CMB lensing and galaxy clustering, between CMB temperature and galaxy clustering, and between CMB temperature and CMB lensing.
Ref.~\cite{Chudaykin:2025gdn} has already shown that the cross-angular power spectrum between the \textit{Planck} PR4 temperature and lensing maps ($C_\ell^{T\kappa}$) imposes a strong constraint on the branding parameter $\alpha_B$.
In addition, it has already been demonstrated that $C_\ell^{T g}$ provides an independent test of general relativity and allows strong constraints to be set on modified gravity scenarios~\cite{Stolzner:2017ged,Renk:2017rzu,Zhao:2025pyo,Dong:2024vxo}. In particular, using pre-DESI and \textit{Planck} 2015 data~\cite{Stolzner:2017ged}, Ref.~\cite{Seraille:2024beb} imposed significant constraints on the EFTofDE parameters.
Concerning $C_\ell^{gg}$ and $C_\ell^{\kappa g}$, they are also highly sensitive to $\alpha_B$ and $\alpha_M$~\cite{Zhao:2008bn,Martinelli:2010wn,Euclid:2023rjj,Euclid:2025tpw}, and can also put strong constraints on the background parameters, especially $\Omega_m$, $h$, $w_0$, $w_a$, and $\sigma_8$~\cite{Reeves:2025xau,DAmico:2025zui,Reeves:2025axp,deBelsunce:2025qku,Kim:2024dmg,Zhang:2024mmg}. A recent study~\cite{Sabogal:2025jbo} has also shown that analyses based on $C_\ell^{gg}$ and $C_\ell^{\kappa g}$ can exhibit a preference for evolving dark energy, even without incorporating primary CMB data.}

In parallel, the effective field theory of large-scale structure (EFTofLSS)~\cite{Carrasco:2012cv,Angulo:2015eqa,Senatore:2014eva,Baumann:2010tm,Perko:2016puo,Fujita:2016dne,Senatore:2014vja,Desjacques:2016bnm,Assassi:2014fva,Vlah:2015sea,Zhang:2021yna} provides a framework for extracting maximum information from the galaxy overdensity field.
It provides an accurate modeling of the 3D galaxy power spectrum and bispectrum~\cite{DAmico:2022ukl}, incorporating the effects of redshift-space distortions~\cite{Senatore:2014vja,Lewandowski:2015ziq}, galaxy bias~\cite{Desjacques:2016bnm,Mirbabayi:2014zca,Assassi:2014fva}, and IR resummation~\cite{Senatore:2014via,Lewandowski:2018ywf,Senatore:2017pbn}, thereby allowing robust cosmological parameter inference~\cite{DAmico:2020kxu,Nishimichi:2020tvu,DAmico:2022osl}.
In particular, the one-loop EFTofLSS prediction of the galaxy power spectrum  have made possible the determination of the $\Lambda$CDM parameters from the full-shape analysis of (e)BOSS data~\cite{BOSS:2016wmc} at precision higher than that from conventional BAO and redshift-space distortions (BAO/$f \sigma_8$) analyses (see Ref.~\cite{Simon:2022csv}), and even comparable to that of CMB experiments. 
Additionally, the inclusion of the one-loop bispectrum yields an additional $\sim30\%$ gain in constraining power on the dark energy equation of state~\cite{Spaar:2023his,Lu:2025gki}.
Based on the equivalence principle~\cite{Creminelli:2013mca,Kehagias:2013yd,DAmico:2021rdb,Peloso:2013zw}, the EFTofLSS can further consistently capture the impact of modified gravity~\cite{Piga:2022mge,Taule:2024bot} and dark energy~\cite{Lu:2025gki,Bose:2018orj,DAmico:2020tty,Lewandowski:2016yce,Silva:2025twg} from the full shape  of the galaxy power spectrum and bispectrum.

The main goal of this work is therefore to comprehensively assess the constraining power of the current measurements of tomographic angular power spectra (derived from photometric surveys and CMB data) and the 3D galaxy power spectrum and bispectrum on the Chevalier-Polarski-Linder (CPL) parametrization of the DE equation of state, with and without the EFTofDE framework to describe the DE perturbations.
To do so, we use (i) the EFTofLSS full-shape analysis of the power
spectrum and bispectrum of BOSS Luminous Red
Galaxies, (ii) the angular power spectrum $C_\ell^{T \kappa}$ from the cross correlation between the \textit{Planck} PR4 temperature and lensing maps, and (iii) the auto angular galaxy power spectra $C_\ell^{gg}$ from DESI luminous red galaxies, together with the cross-angular power spectra $C_\ell^{\kappa g}$ and $C_\ell^{T g}$ from the cross-correlation with the \textit{Planck} PR4 lensing and temperature maps.

This paper is organized as follows. In Sec.~\ref{sec:theory}, we describe the theoretical framework by introducing the EFTofDE setup used in this work, while in Sec.~\ref{sec:analysis_setup} we detail the analysis setup and the modeling of the 3D galaxy power spectrum and angular power spectra mentioned above.
In Sec.~\ref{sec:result}, we present and discuss the cosmological constraints obtained on the CPL parametrization and on the EFTofDE parameters, before concluding in Sec.~\ref{sec:conclusion}.

\section{Theoretical framework: Dark Energy models}\label{sec:theory}

At the background level, DE can be phenomenologically described as an effective fluid characterized by an equation of state $ w(a) \equiv p_{\rm DE}(a)/\rho_{\rm DE}(a)$.
In this work, we adopt the commonly used CPL parametrization~\cite{Chevallier:2000qy,Linder:2002et}, where $ w(a) = w_0 + w_a(1 - a)$.
To specify the DE perturbations, we consider two different frameworks: (i) a phenomenological fluid parametrization called \textit{parametrized post-Friedmann}, and (ii) a physically motivated effective field theory (EFT) approach that captures broad classes of single scalar field and modified gravity models (in this work, we focus on Horndeski's theories).

\subsection{Dark energy as a fluid}

The \textit{Parametrized Post-Friedmann} (PPF) framework~\cite{Hu:2008zd,Fang:2008sn,Dakin:2019vnj} is a phenomenological description of the DE perturbations that is commonly adopted to safely allow a phantom crossing.
Since dark energy is not directly coupled to matter, its perturbations affect structure formation only through the metric potentials in the Poisson equation.
In the conformal Newtonian gauge (denoted ``N''), it reads
\begin{equation}
\begin{aligned}
    k^2 \phi =& -4\pi G a^2 \left( \delta\rho_{\mathrm{tot}}^{\mathrm{N}} 
    - 3\mathcal{H}(\rho_{\mathrm{tot}} + p_{\mathrm{tot}})\frac{\theta_{\mathrm{tot}}^{\mathrm{N}}}{k^2} \right) \\
    =& k^2 \Gamma - 4\pi G a^2 \left( \delta\rho_{\mathrm{t}}^{\mathrm{N}} 
    - 3\mathcal{H}(\rho_{\mathrm{t}} + p_{\mathrm{t}})\frac{\theta_{\mathrm{t}}^{\mathrm{N}}}{k^2} \right) \, .
\end{aligned}
\end{equation}
{In the second line, we isolate the contribution of DE perturbations (expressed in the DE rest frame) as 
$ k^2 \Gamma \equiv -4\pi G a^2 \delta\rho_{\mathrm{DE}}^{\mathrm{rest}} $,
while the subscript ``t'' denotes all species other than dark energy.}

{The evolution equation of $\Gamma$ is obtained from an interpolation between the superhorizon and subhorizon limits:}
\begin{equation}
    \dot{\Gamma} = H \left[
    S\left(1+\frac{c_\Gamma^2 k^2}{\mathcal{H}^2}\right)^{-1}
    - \Gamma \left(1+\frac{c_\Gamma^2 k^2}{\mathcal{H}^2}\right)
    \right],
\end{equation}
where \( c_\Gamma \) is an effective sound speed controlling the transition between large and small scales, and $S$ is defined as 
\begin{equation}
    S\equiv\frac{4\pi Ga^2}{\mathcal H^2}(\rho_{\rm DE}+p_{\rm DE})\frac{\theta^{\rm N}_t}{k^2} \, .
\end{equation}
The CPL + PPF parametrization therefore provides a flexible and simple phenomenological framework for studying deviations from a cosmological constant at both the background and linear perturbation levels.
{Throughout this work, we fix the fluid sound speed to unity $c_s=1$ and set the effective sound speed $c_\Gamma=0.4 \cdot c_s$, following the convention of Ref.~\cite{Fang:2008sn}.}

\subsection{EFTofDE framework}

\begin{figure}
    \centering
    \includegraphics[width=1.\linewidth]{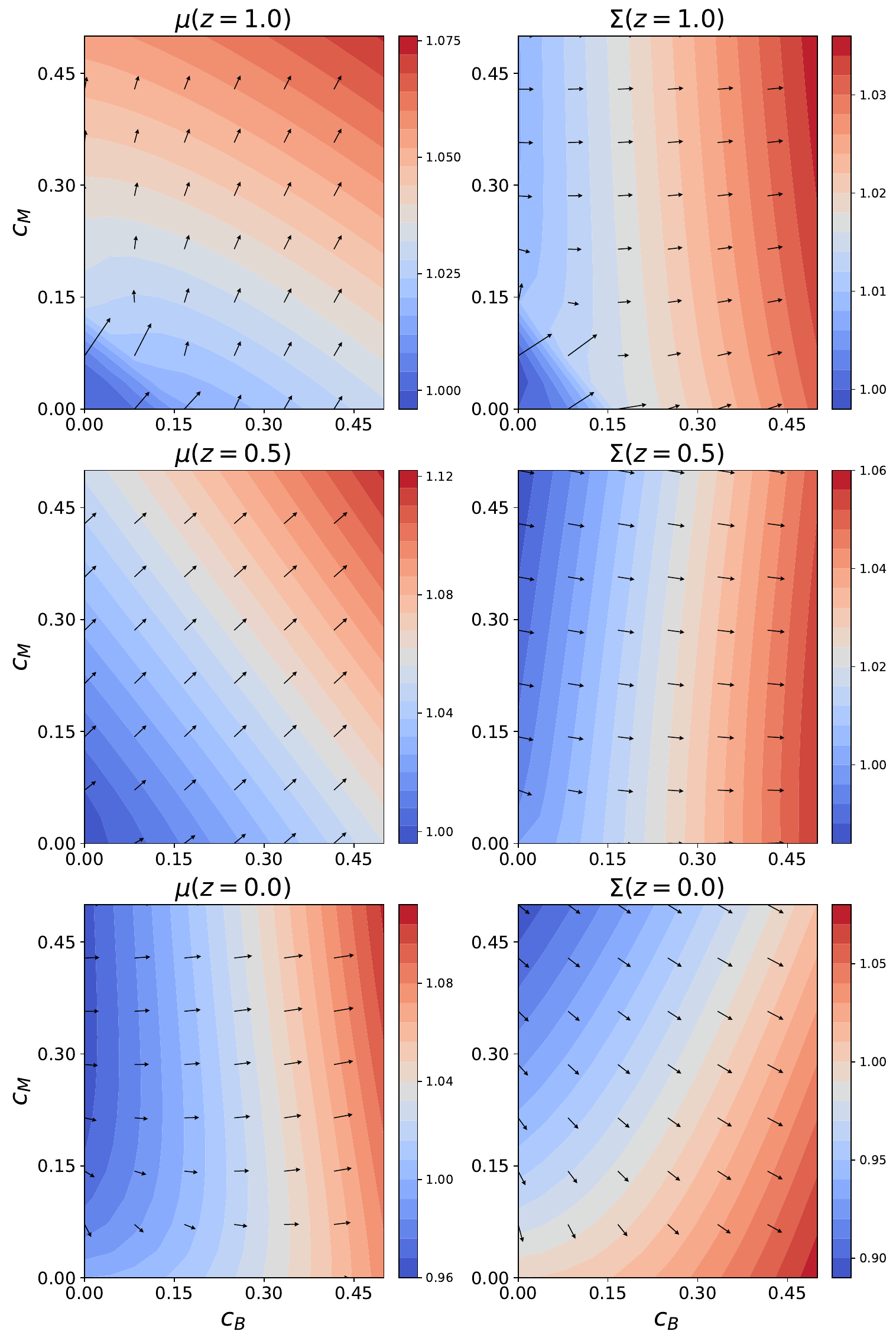}
    \caption{{The response of $\mu$ and $\Sigma$ to variations in $c_M$ and $c_B$ at redshifts $z = 1.0$, $0.5$, and $0.0$. The arrows indicate the gradient direction: $\mu$ is more sensitive to $c_M$ at high redshift and to $c_B$ at low redshift, while $\Sigma$ shows the opposite trend. The cosmology is fixed to the best-fit values of the ``All'' analysis obtained in Sec.~\ref{sec:res_all}}.}
    \label{fig:vary_mu_sigma_against_cBcM}
\end{figure}

EFTofDE provides a physically motivated model-independent description of the DE perturbations around an FLRW background for theories with a single scalar degree of freedom~\cite{Cheung:2007st,Zumalacarregui:2016pph}. Working in unitary gauge and including all operators {that are} invariant under time-dependent spatial diffeomorphisms, the quadratic action for linear scalar and tensor perturbations can be parametrized by a small set of time-dependent functions. Restricting to Horndeski-type theories~\cite{Zumalacarregui:2016pph,Horndeski:1974wa,Nicolis:2008in,Deffayet:2011gz,Kobayashi:2011nu}, whose equations of motion contain at most second-order derivatives, it is customary to introduce four dimensionless EFT functions:
$\alpha_K$, $\alpha_B$, $\alpha_M$, and $\alpha_T$,
which control the kinetic term of the scalar field, the braiding between the kinetic terms of the scalar and the metric, the running of the Planck mass, and tensor speed excess, respectively. 
We note that the parameter $\alpha_M$ is related to the effective Planck mass via
\begin{equation}
\alpha_M\equiv\frac{d\ln M_*^2}{d\ln a}.
\end{equation}
In principle, the EFT parameters can have arbitrary time dependence. In this work, we consider that $\alpha_i(a)=c_i \cdot \Omega_{\rm DE}(a)$~\cite{Bellini:2014fua,Linder:2015rcz,Linder:2016wqw,Taule:2024bot}, allowing us to recover general relativity at high redshifts.
{We note that this choice is not unique and that our cosmological results depend on this particular parametrization. However, this choice is able to capture a wide variety of Horndeski models~\cite{Pujolas:2011he,Barreira:2014jha,Bellini:2015oua}.}

A minimal set of stability {conditions} for linear scalar perturbations is the absence of ghost and gradient instabilities. In terms of the EFT functions, {we can respectively express those conditions as}~\cite{Zumalacarregui:2016pph,Cataneo:2024uox}
\begin{align} \label{eq:stability}
D_{\rm kin}\equiv\alpha_K+\tfrac{3}{2}\alpha_B^2>0 \, , \quad c_s^2>0 \, ,
\end{align}
where $c_s$ corresponds to the scalar propagation speed, which depends on the background expansion history, the $\alpha$-functions, and the matter content as~\cite{Ishak:2024jhs}
\begin{align}\label{eq:stable_condition}
c_s^2=\frac{1}{D_{\rm kin}}\Big[&(2-\alpha_B)\Big(-\frac{H'}{aH^2}+\tfrac{1}{2}\alpha_B+\alpha_M\Big) \nonumber \\
&-\frac{3(\rho_t+p_t)}{H^2M_*^2}+\frac{a_B'}{aH}\Big]\,,
\end{align}
where {$' = d/d\tau$} denotes derivative with respect to the conformal time.
Let us note than once the $\alpha$-functions are set, some parts of the $\{ w_0, \, w_a \}$ parameter space allowed in the CPL + PPF framework become theoretically forbidden due to the EFTofDE stability conditions~\cite{Zumalacarregui:2016pph,Cataneo:2024uox}.

Restricting to the quasi-static approximation, where we can neglect the time evolution of the scalar degree of freedom~\cite{Cusin:2017mzw,Cusin:2017wjg,Bose:2018orj}, 
the EFT functions modify linear perturbations  through two effective functions, namely, $\mu(k,a)$, which rescales the Poisson equation and {modifies the matter} clustering, and $\Sigma(k,a)$, which rescales the Weyl potential and controls the photon trajectories~\cite{Sakr:2021ylx,Pogosian:2016pwr}:
\begin{align}
    &k^2\Psi=-4\pi Ga^2\mu(k,a)\sum_i\rho_i\Delta_i\,, \label{eq:poisson}\\
    &k^2(\Psi+\Phi)=-8\pi Ga^2\Sigma(k,a)\sum_i\rho_i\Delta_i\,, \label{eq:weyl}
\end{align}
where $\rho_i$ and $\Delta_i$ denote the density and the gauge-invariant overdensity fields of the species $i$, respectively.
Within the EFTofDE framework, the functions $\mu(k,a)$ and $\Sigma(k,a)$ can be written in terms of $\{M_*,\, \alpha_i,\, c_s^2\}$ as~\cite{Zumalacarregui:2016pph,Ishak:2024jhs,Bellini:2014fua}
\begin{align}
&\mu(a)=\frac{1}{M_*^2}\left[1+\frac{2(\alpha_M+\frac{1}{2}\alpha_B)^2}{c_s^2(\alpha_k+\frac{3}{2}\alpha_B^2)}\right]\,,\label{eq:mu}\\
&\Sigma(a)=\frac{1}{M_*^2}\left[1+\frac{(\alpha_M+\frac{1}{2}\alpha_B)(\alpha_M+\alpha_B)}{c_s^2(\alpha_k+\frac{3}{2}\alpha_B^2)}\right]\,. \label{eq:sigma}
\end{align}
{These functions characterize deviations from general relativity (GR) by modifying the relations between the metric potentials and the matter perturbation field, and are therefore directly constrained by large-scale structure and CMB observations.
Note that in this work $\mu$ and $\Sigma$ are only functions of time. 
In addition, we set $c_K = 1$ because within the quasi-static limit studied in this work, the cosmological data are insensitive to this parameter~\cite{Copeland:2018yuh,Bellini:2015xja,Zumalacarregui:2016pph,Alonso:2016suf,Gleyzes:2015rua,Chudaykin:2024gol}. We also set $c_T = 0$ since the simultaneous detection of GW170817 and its electromagnetic counterpart~\cite{LIGOScientific:2017zic} set a strong constraint on this parameter with $\alpha_T \lesssim 10^{-15}$~\cite{Ezquiaga:2017ekz,Baker:2017hug,Kreisch:2017uet,Creminelli:2017sry,Copeland:2018yuh}.
Therefore, in this paper, we aim to constrain the theoretical parameter space $\{ c_B, \, c_M \}$.
Fig.~\ref{fig:vary_mu_sigma_against_cBcM} illustrates the dependence of $\mu(a)$ and $\Sigma(a)$ on $c_B$ and $c_M$ at different redshifts. In particular, at high redshifts $\mu(a)$ mainly responds to variations in $c_M$, whereas at low redshifts it is more sensitive to $c_B$, while $\Sigma(a)$ exhibits the opposite trend. 
Therefore, the dependence of $\mu(a)$ and $\Sigma(a)$ on $c_B$ and $c_M$ depends strongly on the redshift, which is a key feature that motivates a multiprobe analysis spanning a wide redshift range. }

\section{Analysis setup}\label{sec:analysis_setup}

\subsection{Datasets and inference framework}

In this paper, we perform Markov chain Monte Carlo (MCMC) analyses using the Metropolis-Hastings algorithm from the \texttt{MontePython-v3}\footnote{\url{https://github.com/brinckmann/montepython_public}} code \cite{Audren:2012wb,Brinckmann:2018cvx} interfaced with {\texttt{hiclass}},\footnote{\url{https://miguelzuma.github.io/hi_class_public/}} a modified version of {\texttt{CLASS}}\footnote{\url{https://lesgourg.github.io/class_public/class.html}}  \cite{Blas:2011rf} including EFTofDE cosmologies (see Refs.~\cite{Raveri:2014cka,Bellini:2017avd} for alternative codes).
In the following, we describe the various likelihoods used in our analysis.

\subsubsection{Datasets}

Our baseline dataset is made up of the following likelihoods:
\begin{itemize}
    \item \textbf{Planck PR4}: The high-$\ell$ TT, TE, EE, and low-$\ell$ EE CMB power spectrum data (dubbed $C_\ell^{TT}$, $C_\ell^{EE}$, and $C_\ell^{TE}$ in the following) from the \textit{Planck} Public Release 4 (PR4) analyzed with the \texttt{NPIPE} processing pipeline~\cite{Planck:2020olo}, respectively described using the \texttt{Hillipop} and \texttt{LoLLiPoP} likelihoods \cite{Tristram:2020wbi, Tristram:2023haj}. Similar to \textit{Planck} Public Release 3 (PR3), we also include the low-$\ell$ TT likelihood using \texttt{Commander}~\cite{Planck:2019nip}.
    \item \textbf{Lensing:} The CMB gravitational lensing spectrum $C_\ell^{\kappa \kappa}$ from \textit{Planck} PR4 \cite{Carron:2022eyg} analyzed in the multipole range $8 < L < 400$.
    Note that we have adapted the public \texttt{Cobaya} likelihood\footnote{\url{https://github.com/carronj/planck_PR4_lensing}} to \texttt{Montepython-v3}.
    \item \textbf{DESI DR2 BAO:} The DESI DR2 BAO data from Refs.~\cite{DESI:2025zgx,DESI:2025zpo} include BAO from bright galaxies ($0.1<z<0.4$), luminous red galaxies ($0.4<z<1.1$), emission line galaxies ($0.8<z<1.6$), quasars ($0.8<z<2.1$), and Lyman$-\alpha$ forest quasars ($1.8<z<4.2$). 
    \item \textbf{DES Y5:}  DES year 5\footnote{In the recent update of the DES supernova analysis~\cite{DES:2025sig}, the evidence for evolving dark energy has been reduced from $4.2\sigma$ to $3.2\sigma$.} catalog of uncalibrated luminosity distance of type Ia supernovae (SNeIa) in the range $0.10 < z < 1.13$, combined with an external sample of SNeIa at low redshifts from $0.024<z<0.10$~\cite{DES:2024tys}. We have adapted the public \texttt{Cobaya}~\cite{Torrado:2020dgo} likelihood to \texttt{Montepython-v3} in Ref.~\cite{Lu:2025gki}.
\end{itemize}
The main purpose of this work is to combine this baseline dataset with the following likelihoods in order to assess their impact on the constraints on the background expansion and gravitational properties:
\begin{itemize}
    \item \textbf{ISWL:} The angular power spectrum $C_\ell^{T \kappa}$ from the cross-correlation between the \textit{Planck} PR4 temperature map \cite{Planck:2020olo} and the \textit{Planck} PR4 lensing map \cite{Carron:2022eyg} from Ref.~\cite{Carron:2022eum} analyzed in the multipole range $2 < L < 100$.  
    Note that we have adapted the public \texttt{Cobaya} likelihood\footnote{\url{https://github.com/carronj/planck_PR4_lensing}} to \texttt{Montepython-v3}.\footnote{{We have checked that our \texttt{Montepython-v3} likelihood is consistent with the original \texttt{Cobaya} one up $0.2\sigma$.}}
    \item \textbf{EFTBOSS:} The full-shape analysis of the power spectrum and bispectrum of BOSS luminous red galaxies (LRGs)~\cite{BOSS:2016wmc}. To model those observables, we use the effective field theory of large-scale structure (EFTofLSS) up to one loop~\cite{Carrasco:2012cv,Angulo:2015eqa,Senatore:2014eva,Baumann:2010tm,Perko:2016puo,Fujita:2016dne,Senatore:2014vja,Desjacques:2016bnm,Assassi:2014fva,Vlah:2015sea,Zhang:2021yna} (see below for more details).
    The power spectrum and bispectrum are measured in Refs.~\cite{DAmico:2022osl,DAmico:2022ukl,Spaar:2023his,Ivanov:2021kcd} from the BOSS catalogs DR12 (v5) combined CMASS-LOWZ\footnote{Publicly available at \url{https://data.sdss.org/sas/dr12/boss/lss/}}~\cite{BOSS:2015ewx}, and are divided in two redshift bins (and four sky cuts): low-$z$, $0.2<z<0.43 \  (z_{\rm eff}=0.32)$, and high-$z$, $0.43<z<0.7  \ (z_{\rm eff}=0.57)$, with north and south Galactic skies for each. The covariance, including the correlation between power spectrum and bispectrum, is measured through 2048 Patchy mocks~\cite{Kitaura:2015uqa}, while the window function is measured from \texttt{fkpwin}\footnote{\url{https://github.com/pierrexyz/fkpwin}}~\cite{Beutler:2018vpe}. 
    \item \textbf{DESI$C_\ell$:} 
    {The auto-angular galaxy power spectra $C_\ell^{gg}$ from a photometric sample of luminous red galaxies (LRGs) of the DESI Legacy Imaging Survey DR9~\cite{Zhou:2023gji,DESI:2022gle}, together with the cross-angular power spectra $C_\ell^{\kappa g}$ from the cross-correlation between the same LRG photometric sample and the \textit{Planck} PR4 lensing map~\cite{Carron:2022eyg}.} 
    {The data and covariances, analyzed in the multipole range $20\leq L \leq 243$ and separated into four photometric redshift bins {(with the effective redshifts $z_{\rm eff} = \{ 0.470, \, 0.625, \, 0.785, \, 0.914 \}$)}, are determined in Refs.~\cite{Sailer:2024jrx,Kim:2024dmg} using the \texttt{NaMaster} code\footnote{\url{https://github.com/LSSTDESC/NaMaster}}~\cite{Alonso:2018jzx}. Our \texttt{MontePyhton-v3} likelihood--inspired from the \texttt{MaPar} likelihood\footnote{\url{https://github.com/NoahSailer/MaPar/tree/main}} written in \texttt{Cobaya}~\cite{Sailer:2024jrx,Kim:2024dmg}--does not consider the Limber approximation as in Refs.~\cite{Sailer:2024jrx,Kim:2024dmg}, but uses \texttt{Swift$C_\ell$}\footnote{\url{https://cosmo-gitlab.phys.ethz.ch/cosmo_public/swiftcl}}~\cite{Reymond:2025ixl} instead, which is an accurate and differentiable JAX-based code computing the angular galaxy power spectrum beyond the Limber approximation (see below). In addition, we also measure the cross-angular power spectra $C_\ell^{T g}$ from the cross-correlation between the DESI DR9 LRG photometric samples and the \textit{Planck} PR4 temperature map~\cite{Planck:2020olo}, where we use \texttt{NaMaster} to determine the data and covariances and \texttt{Swift$C_\ell$} to compute the theoretical prediction. We refer the reader to App.~\ref{appendix:mapar} for more details on the determination of the $C_\ell^{T g}$ data as well as on the difference between the Limber approximation and \texttt{Swift$C_\ell$}. The data used in this likelihood ($C_\ell^{gg}$, $C_\ell^{\kappa g}$ and $C_\ell^{T g}$) and the theoretical predictions are shown in Fig.~\ref{fig:namaster_cltg} of this appendix.} The validation of our implementation is detailed in Sec.~\ref{appendix:mapar}, while the likelihood is publicly available on GitHub \github{https://github.com/GreenPlanck/DESICl}.
\end{itemize} 

\subsubsection{Correlations between datasets}\label{sec:data_corr}

{In our analyses, we use the $C_\ell^{TT}$, $C_\ell^{EE}$, $C_\ell^{\kappa\kappa}$ and $C_\ell^{gg}$ auto-angular power spectra, while we use the $C_\ell^{TE}$ (from \textit{Planck} PR4), $C_\ell^{T\kappa}$ (from ISWL), $C_\ell^{Tg}$ (from DESI$C_\ell$), and $C_\ell^{\kappa g}$ (from DESI$C_\ell$) cross-angular power spectra.
The objective of this paper is therefore to perform a multiprobe analysis to exploit as much information as possible from several cosmological probes and their cross-correlations. However, caution must be taken when performing such an analysis to avoid double counting information. Below, we list how we take into account the various correlations in our analysis:}
\begin{itemize}
    \item {The correlations between the \textit{Planck} ($C_\ell^{TT}$, $C_\ell^{EE}$ and $C_\ell^{TE}$), lensing ($C_\ell^{\kappa\kappa}$) and ISWL ($C_\ell^{T\kappa}$) likelihoods are included in our analysis following Refs.~\cite{Carron:2022eyg,Carron:2022eum}.}

    \item There is a non-negligible covariance between $C_\ell^{\kappa\kappa}$ (from the lensing likelihood) and $C_{\ell}^{\kappa g}$ (from the DESI$C_\ell$ likelihood). For simplicity, when combining $C_{\ell}^{\kappa g}$ with $C_\ell^{\kappa\kappa}$, we remove the overlapping multipoles by excluding $\ell<243$ for $C_\ell^{\kappa\kappa}$ (see, \textit{e.g.}, Ref.~\cite{Reeves:2025xau} for an estimation of such a covariance).

    \item {There are also correlations between $C_\ell^{\kappa g}$ (DESI$C_\ell$) and $C_\ell^{\kappa T}$ (ISWL), as well as between $C_\ell^{T g}$ (DESI$C_\ell$) and $C_\ell^{TT - TE}$/$C_\ell^{T \kappa}$ (\textit{Planck}/ISWL).}
    {However, using a Gaussian approximation to compute the full covariance}~\cite{Zhang:2021wzo,Taylor:2022rgy},
    \begin{align}
        {\rm Cov}_{LL'}^{\rm AB,CD} =
    &\frac{\delta_{LL'}}{(2\ell_{L'}+1)\,\triangle \ell\, f_{\rm sky}} \nonumber \\
    &\times \Big[
    {C_L^{\rm AC}C_L^{\rm BD}}
    + {C_L^{\rm AD}C_L^{\rm BC}}
    \Big]\,,
    \end{align}
    {the correlation coefficient $\rho(\ell)=\frac{{\rm Cov}(C_\ell^{T x},C_\ell^{AB})}{\sigma(C_\ell^{Tx})\sigma(C_\ell^{AB})}$ is found to be below $8\%$ for these cross-correlations and is therefore neglected in this work.}

    \item {As in Ref.~\cite{Maus:2025rvz}, we neglect the correlation between photometric and spectroscopic data given that they probe different cosmological volumes (see Ref.~\cite{Taylor:2022rgy} for more details), implying that we can safely combine the DESI$C_\ell$ likelihood with the DESI DR2 BAO and EFTBOSS likelihoods.}

    \item Finally, in our analysis, we combine DESI DR2 BAO with EFTBOSS. To mitigate the correlation between these two likelihoods, we use the sound horizon $r_s$ information from the BAO analysis of DESI, while we marginalize it in the EFTBOSS full-shape analysis, as done in Ref.~\cite{Lu:2025gki}. To do so, we follow Refs.~\cite{Farren:2021grl, Smith:2022iax}, splitting the linear power spectrum into its broadband shape and its  wiggle component, before marginalizing over a new scaling parameter, dubbed $\alpha_{r_s}$.
    In DESI DR1, approximately $27\%$ of BGS and LRG1 galaxies and $9\%$ of LRG2 galaxies were already observed by SDSS~\cite{DESI:2024aax}, which induces a $\sim10\%$ correlation between the two datasets, reduced to less than $5\%$ when marginalizing over the sound horizon $r_s$~\cite{Lu:2025gki}. This overlap increases in DESI DR2~\cite{DESI:2025zgx}, and we evaluate that the correlation is at most $10\%$\footnote{We note that the DESI BGS and LRG samples together constitute approximately $28\%$ of the total DESI BAO volume (see Tab.~III in Ref.~\cite{DESI:2025zgx}). As a conservative estimate, we assume that roughly half of the galaxies in the DESI BGS and LRG samples were already observed by BOSS. Assuming further that BOSS covers half the data volume of DESI, and noting that the $r_s$-marginalization procedure removes $55\%$ of the correlation~\cite{Lu:2025gki}, we obtain an upper bound on the residual correlation of $\rho =  \frac{0.5 \times V_{\rm BGS,LRG}}{\sqrt{V_{\rm DESI} V_{\rm BOSS}}}\times45\% \sim 10\%$, where $0.5 \times V_{\rm BGS,LRG}$ corresponds to the overlapping volume between BOSS and DESI.} after the marginalization over the sound horizon. 
    To further support our results, each time we combine DESI DR2 BAO with EFTBOSS, we reproduce the same analysis by replacing DESI DR2 BAO with ext-BAO, a likelihood--that can be safely combined with BOSS--including BAO measurements from 6dFGS at $z=0.106$~\cite{Beutler:2011hx}, SDSS DR7 at $z=0.15$~\cite{Ross:2014qpa}, as well as eBOSS DR14 Ly-$\alpha$ absorption auto-correlation at $z = 2.34$ and cross-correlation with quasars at $z = 2.35$~\cite{eBOSS:2019ytm,Cuceu:2019for}. 

\end{itemize}

\subsubsection{Analysis settings}

For all runs performed, we impose large flat priors on the $\Lambda$CDM parameters $\{\omega_b, \,\omega_{\rm cdm},\, H_0,\, A_s,\, n_s,\, \tau_{\rm reio}\}$, respectively corresponding to the dimensionless baryon energy density, the dimensionless cold dark matter energy density, the current Hubble parameter, the variance of the primordial power spectrum at $k = 0.05 \, {\rm Mpc}^{-1}$, the primordial power spectrum tilt, and the optical depth to reionization. We also impose large flat priors on the DE background parameters $\{ w_0,w_a \}$ and on the EFTofDE pertubations parameters $\{ c_B, c_B \}$.

Note that throughout this paper we use the \textit{Planck} neutrino treatment convention, including two massless and one massive species with $m_{\nu} = 0.06$ eV \cite{Planck:2018vyg}.
We consider that our chains have converged when the Gelman-Rubin criterion $R-1<0.05$.
We acknowledge the use of \texttt{Procoli}~\cite{Karwal:2024qpt} to extract the best-fit parameters and the best-fit $\chi^2$ as well as the use of \texttt{GetDist}~\cite{Lewis:2019xzd} to extract the probability density functions and produce our plots.

\subsection{Modeling and theoretical predictions}

In this section, we describe the modeling and theoretical predictions associated with the three likelihoods we are adding to our baseline dataset, namely, EFTBOSS (in Sec.~\ref{sec:EFTBOSSmodelling}), ISWL, and DESI$C_\ell$ (in Sec.~\ref{sec:angularspectramodelling}).

\subsubsection{Galaxy power spectrum and the EFTofLSS} \label{sec:EFTBOSSmodelling}
Modeling the nonlinear evolution of large-scale structure and connecting the galaxy overdensity field to the underlying dark matter overdensity field is challenging. 
In this paper, we adopt the effective field theory of large-scale structure (EFTofLSS) framework allowing us to organize the expansions in fluctuations and derivatives of the density
and velocity fields of the observed tracers at long wavelengths, which allows us to compute the galaxy power spectrum and bispectrum in redshift space up to one loop.\footnote{The first formulation of the EFTofLSS was carried out in Eulerian space in Refs.~\cite{Carrasco:2012cv,Baumann:2010tm} and in Lagrangian space in \cite{Porto:2013qua}. Once this theoretical framework was established, many efforts were made to improve this theory and make it predictive, such as the understanding of renormalization \cite{Pajer:2013jj, Abolhasani:2015mra}, the IR-resummation of the long displacement fields \cite{Senatore:2014vja, Baldauf:2015xfa, Senatore:2014via, Senatore:2017pbn, Lewandowski:2018ywf, Blas:2016sfa}, and the computation of the two-loop matter power spectrum \cite{Carrasco:2013sva, Carrasco:2013mua}}
Note that following previous studies (see, \textit{e.g}, Ref.~\cite{Lu:2025gki}) showing that the Einstein-de Sitter (EdS) approximation is sufficient for BOSS data, we adopt it here, with the time dependence of density perturbations given by $\delta^{(n)}(a) \propto D_+^n(a)$.

To determine the full-modeling information of BOSS data (denoted ``EFTBOSS''), we adopt the \texttt{PyBird} EFT likelihood\footnote{\url{https://github.com/pierrexyz/pybird}}~\cite{DAmico:2020kxu}, for which we specify all the important details below:
\begin{itemize}

    \item \textbf{Renormalization scales:} In the redshift-space galaxy power spectrum, there are three renormalization scales~: the nonlinear scale renormalizing the dark matter field $k_{\rm NL}$, the spatial extension of the observed objects renormalizing the spatial derivative expansion $k_{\rm M}$, and the ``dispersion'' scale renormalizing the velocity products appearing in the redshift-space expansion $k_{\rm R}$. According to Refs.~\cite{DAmico:2019fhj,DAmico:2021ymi}, we set $k_{\rm M} = k_{\rm NL} = 0.7 \, h {\rm Mpc}^{-1}$ and $k_{\rm R} = 0.25 \, h {\rm Mpc}^{-1}$. In addition, we set the mean galaxy number density to $\Bar{n}_g = 4 \cdot 10^{-4} \, ({\rm Mpc}/h)^3$.

    \item \textbf{Priors:} For all the EFT parameters, we impose a Gaussian prior centered on 0 with a width of 2, except for the linear bias parameter $b_1$ where we impose a log-normal prior, $\log b_1\sim \mathcal N(0.8,0.8944)$ (see Refs.~\cite{DAmico:2022osl,DAmico:2022ukl} for all the details on the EFT parameters and the associated priors).
    Following Ref.~\cite{DAmico:2020kxu}, we analytically marginalize all the EFT parameters that enter linearly in the theory, while we vary the other parameters, namely, $\{b_1, \, b_2, \, b_5 \}$ (see Ref.~\cite{DAmico:2022ukl}), allowing us to considerably reduce the parameter space probed by the MCMC, going from 41 EFT parameters per sky cut to only 3. Finally, we consider correlations between the sky cuts--thanks to a multivariate Gaussian prior--by imposing the EFT parameters to vary by at most $10 \%$ between north and south hemispheres, and by  at most $20 \%$ between the low-$z$ and high-$z$ samples (see Refs.~\cite{DAmico:2022osl,Spaar:2023his}).
    
    \item \textbf{Cutoff scales:} For both the power spectrum and the bispectrum, we consider $ k \in [0.01, 0.20] \, h {\rm Mpc}^{-1}$ for the low-$z$ sample and $k \in [0.01, 0.23] \, h {\rm Mpc}^{-1}$ for the high-$z$ sample (see Refs.~\cite{Colas:2019ret,DAmico:2020kxu,DAmico:2022osl}).

    \item \textbf{Observational effects:} Our analysis includes several observational modelings (see, \textit{e.g.}, Ref.~\cite{DAmico:2019fhj}), such as the window functions from Ref.~\cite{Beutler:2018vpe} (see App.~A of Ref.~\cite{Simon:2022adh} for more details), the binning effect~\cite{DAmico:2022osl}, or the Alcock-Paczynski effect~\cite{Alcock:1979mp}.
\end{itemize}

\subsubsection{Angular power spectra}
\label{sec:angularspectramodelling}

We now turn to the description of the modeling of the ISWL and DESI$C_\ell$ likelihoods. The cross-angular power spectrum between two tracers $i,\, j$ on the 2D full sky is computed from the following line-of-sight integral (see, \textit{e.g.}, Ref.~\cite{Dodelson:2003ft})
\begin{equation}\label{eq:cl}
    C_\ell^{ij}=\frac{2}{\pi}\int dk k^2 \Delta_\ell^i(k)\Delta_\ell^j(k)P(k,\chi,\chi')\,,
\end{equation}
where $P(k,\chi,\chi')$ corresponds to the unequal-time matter power spectrum, and where $\Delta_\ell^i(k)$ denotes the source function for the tracer $i$. The latter takes the form of a time-dependent window function $W^i(\chi)$ which is line-of-sight integrated over the spherical Bessel functions, together with an $\ell$-dependent prefactor $c(\ell)$:
\begin{align}
    \Delta_\ell^i(k)=c(\ell)\int _0^\infty d\chi W^i(\chi)j_\ell^{(n)}(k\chi)\,, \label{eq:delta}
\end{align}
where $(n)$ corresponds the n-th derivative.
In this paper, we do not consider the commonly used Limber approximation~\cite{Limber:1954zz,Kaiser:1991qi,LoVerde:2008re}--reducing the triple integral
in Eq.~\eqref{eq:cl} into a single integral over the comoving distance $\chi$--which is valid only at small scales, but we perform the full integral in Eq.~\eqref{eq:cl} using \texttt{Swift$C_\ell$}~\cite{Reymond:2025ixl} (see also Ref.~\cite{Fang:2019xat} for an alternative code). 
In particular, this code uses an FFTlog decomposition of the $\chi$-dependent part of the integrand, in order to analytically compute the integrals over $\chi$ and $\chi'$ before numerically integrating Eq.~\eqref{eq:cl} over $k$.
We refer the interested reader to Ref.~\cite{Reymond:2025ixl} for further detail.\\

In the ISWL and DESI$C_\ell$ likelihoods, we consider the angular power spectra for the following pairs of fields: $(g,g),(g,\kappa),(g,T),(\kappa,T)$.
Therefore, we need to determine the kernels for the galaxy clustering $\Delta_\ell^g$, the CMB lensing $\Delta_\ell^{\kappa}$, and the CMB temperature $\Delta_\ell^T$ in order to obtain the angular power spectra from Eq.~\ref{eq:cl}. In the following, we describe the theoretical predictions for these observables within the EFTofDE framework depicted in Sec.~\ref{sec:theory}.

\textbf{\textit {Galaxy clustering kernel.}} The galaxy clustering source function $\Delta_\ell^g$ implemented in \texttt{Swift$C_\ell$}~\cite{Reymond:2025ixl}  includes contributions from galaxy fluctuations $\Delta_\ell^{g,D}$ (see, \textit{e.g.}, Ref.~\cite{Dodelson:2003ft}), magnification bias $\Delta_\ell^{g,\mu}$~\cite{Villumsen:1995ar,Moessner:1997qs,DES:2021rex}, and redshift-space distortions $\Delta_\ell^{g,\rm RSD}$~\cite{DES:2021rex}, such that 
\begin{equation}
    \Delta_\ell^g=\Delta_\ell^{g,D}+\Delta_\ell^{g,\mu}+\Delta_\ell^{g,\rm RSD}\,,
\end{equation}
where
\begin{equation}\label{eq:clgg_kernel}
\begin{aligned}
    &\Delta_\ell^{g,D}=b_1(z) \int_0^{\infty }d\chi W^{\delta,g}(\chi)D_+(k,z(\chi))j_\ell(k\chi)\,,\\
    &\Delta_\ell^{g,\mu}=\frac{\ell(\ell+1)}{k^2}(5s_\mu-2)\int_0^\infty \frac{d\chi}{\chi^2}W^{\kappa,g}D_+(k,z(\chi))j_\ell(k\chi)\,,\\
    &\Delta_\ell^{g,\rm RSD}=-\int _0^\infty f_+(z(\chi))W^{\delta,g}(\chi)D_+(k,z(\chi))j_\ell^{(2)}(k\chi)\,.
\end{aligned}
\end{equation}
The full galaxy clustering kernel depends on the galaxy fluctuation ($W^{\delta,g}$) and magnification bias ($W^{\kappa,g}$) window functions, which are respectively given by
\begin{equation}
    \begin{aligned}
    &W^{\delta,g}(\chi)=n(z(\chi))\,,\\
    &W^{\kappa,g}(\chi)=\frac{3\Omega_m H_0^2}{2c^2}\int _\chi^\infty d\chi' n(\chi')\frac{\chi}{a(\chi)}\frac{\chi'-\chi}{\chi}\,,
    \end{aligned}
\end{equation}
where $n(z(\chi))$ corresponds the normalized galaxy redshift distribution.
In this set of equations, $D_+(k,z(\chi)) = \sqrt{P(k,z(\chi))/P(k)}$ corresponds to the growth factor and $f_+=d\ln D_+/d\ln a$ corresponds to the growth rate, which are directly extracted from \texttt{hiclass}.
Unlike the EFTBOSS likelihood, we consider a linear galaxy bias model to determine the galaxy overdensity field in the DESI$C_\ell$ likelihood (see Refs.~\cite{Kaiser:1984sw,Bardeen:1985tr}), where $b_1(z)$--defined as $\delta_g = b_1(z) \delta_m$--corresponds to the linear bias parameter. We vary $b_1$ for each LRG redshift bin.
We also include a shot noise contribution ${\rm SN}_i$ in $C_{\ell, i}^{gg}$ for each redshift bin $i$ (see Ref.~\cite{Sailer:2024jrx}), in such way that $C_{\ell, i}^{gg} \to C_{\ell, i}^{gg} + {\rm SN}_i$.
Additionally, the magnification bias kernel depends on one free parameter, $s_{\mu}$, corresponding to the number count slope.
We note that this contribution is subdominant in our analysis, and that the reconstructed posteriors of $s_\mu$ are largely dominated by the priors on the (linear) scales we are considering.
Finally, unlike Ref.~\cite{Sailer:2024jrx}, we also include a contribution from the redshift-space distortion effect, which becomes important at large scale, as we show in App.~\ref{appendix:mapar} (see also Ref.~\cite{Sailer:2024jrx}).

\textbf{\textit {CMB lensing kernel.}} The source function for the CMB lensing is given by
\begin{equation}\label{eq:clkg_kernel}
    \Delta_\ell^\kappa=\ell(\ell+1)\frac{3\Omega_m H_0^2}{2c^2k^2}\int_0^{\chi_*}\frac{d\chi}{a(\chi)}\frac{\chi_*-\chi}{\chi\chi_*}D_+(k,z(\chi))j_\ell(k\chi)\,,
\end{equation}
where $\chi_*$ corresponds to the comoving distance to the last scattering surface. In this equation, we have incorporated the CMB lensing kernel which is analog to the magnification bias kernel (where  we replace the galaxy redshift distribution by the lens redshift distribution).

\textbf{\textit {CMB temperature kernel.}} The CMB gravitational lensing field and the galaxy overdensity field are correlated with the integrated Sachs-Wolfe effect contribution of the CMB temperature field since the same gravitational potentials are responsible for these three effects.
The source function for the CMB temperature anisotropies is given by~\cite{Nicola:2016eua}:
\begin{align}\label{eq:cltg_kernel}
    \Delta_\ell^{T}= &2T_{\rm CMB} \frac{3\Omega_mH_0^2}{2c^3k^2} \nonumber \\
    & \times \int _0^{\chi_*}d\chi(1-\tilde f_+(z(\chi)))D_+(z(\chi))H(z(\chi))j_\ell(k\chi)\,,
\end{align}
where $T_{\rm CMB}$ is the current CMB temperature, and $\tilde f_+=d\ln (D_+\Sigma)/d\ln a$ corresponds to the modified growth rate {(see Ref.~\cite{Seraille:2024beb})}.
The $\Sigma$ factor, defined in Eq.~\ref{eq:sigma}, arises from the Weyl potential $\Phi+\Psi$, which governs the photon geodesics. 
Note that the growth factor $D_+$ here inherently includes modifications to gravity through the $\mu$ parameter.\\

In this work, we adopt the following settings for the DESI$C_\ell$ likelihood:
\begin{itemize}
    \item  \textbf{Cutoff scales:} We consider the following scale cuts for the angular power spectra across the four redshift bins:
    \begin{align*}
    C_\ell^{gg} &: [20,124], [20,178], [20,243], [20,243],\\
    C_\ell^{\kappa g} &: [20,178], [20,178], [20,243], [20,243],\\
    C_\ell^{Tg} &: [20,178], [20,178], [20,178], [20,178].
\end{align*}
     We discuss in App.~\ref{appendix:mapar} how we determine these scale cuts.
    \item \textbf{Priors:} We impose, for each redshift bin, an uniform prior on $b_1$ and Gaussian priors on $s_\mu$ and ${\rm SN}_i$ (see Refs.~\cite{Sailer:2024jrx,Zhou:2023gji}):
\begin{align*}
    b_1 &\sim \mathcal{U}(1,3) \,,\\
    s_\mu^i &\sim \mathcal{N}(1, 0.1) \,, \\
    {\rm SN}_i &\sim \mathcal{N}({\rm SN}_{i, \rm fid}, 0.3 \, {\rm SN}_{i, \rm fid}) \,,
\end{align*}
where ${\rm SN}_{i, \rm fid}$ corresponds to the shot noise fiducial values for the four redshift bins 
${\rm SN}_{i, \rm fid} = \{4.07 \cdot 10^{-6}, \, 2.25 \cdot  10^{-6}, \, 2.05 \cdot 10^{-6}, \, 2.25 \cdot 10^{-6}\}$.
\end{itemize}
For the details (and validations) about the modeling of the DESI$C_\ell$ likelihood, we refer the interested reader to App.~\ref{appendix:mapar}.

\section{Results}\label{sec:result}

\begin{figure*}[t]
    \centering

    % Column titles
    \makebox[0.24\linewidth][c]{EFTBOSS}%
    \makebox[0.24\linewidth][c]{ISWL}%
    \makebox[0.24\linewidth][c]{DESI$C_\ell$}%
    \makebox[0.24\linewidth][c]{All}\\[0.5ex]  % line break with small spacing

    % First row
    \includegraphics[width=0.24\linewidth]{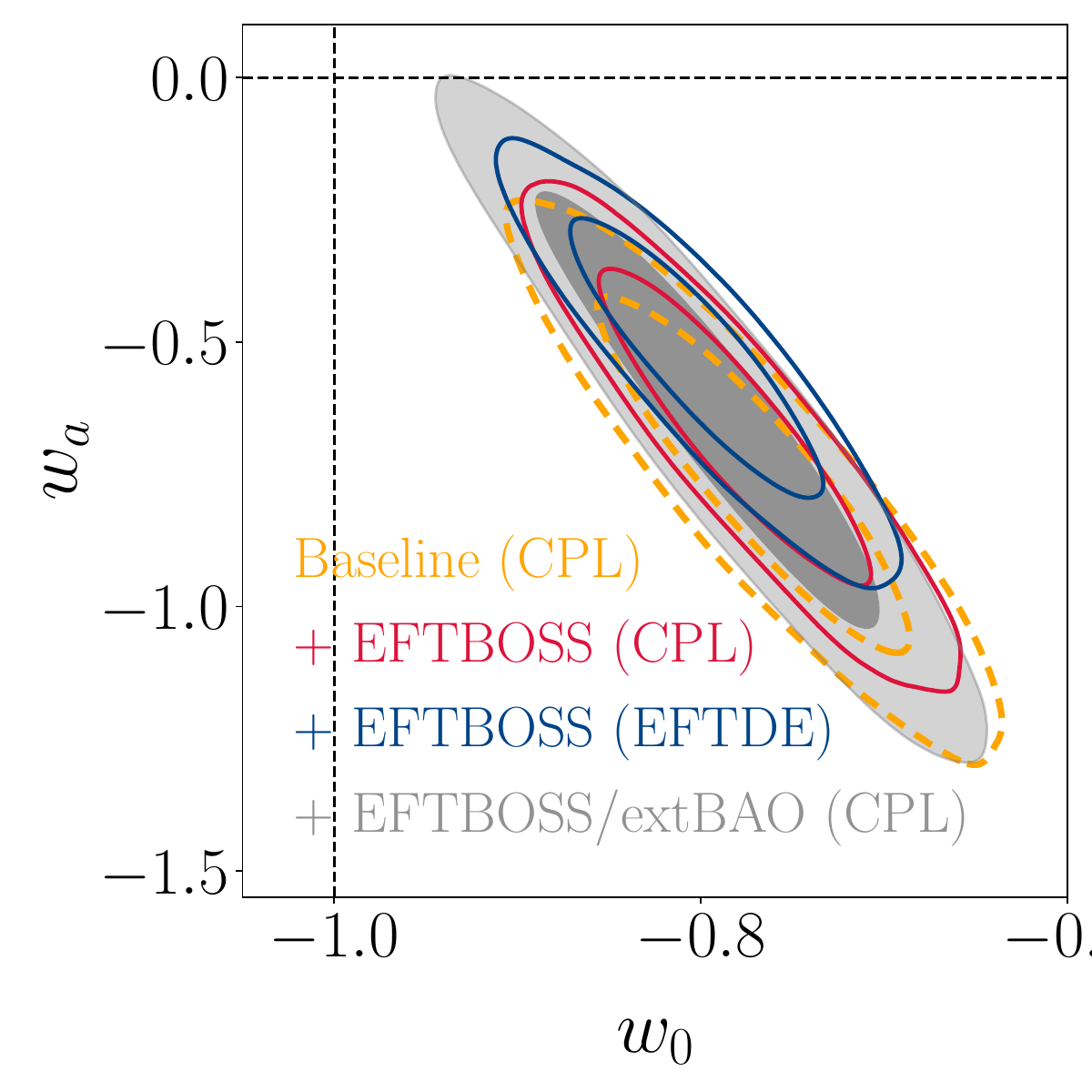}
    \includegraphics[width=0.24\linewidth]{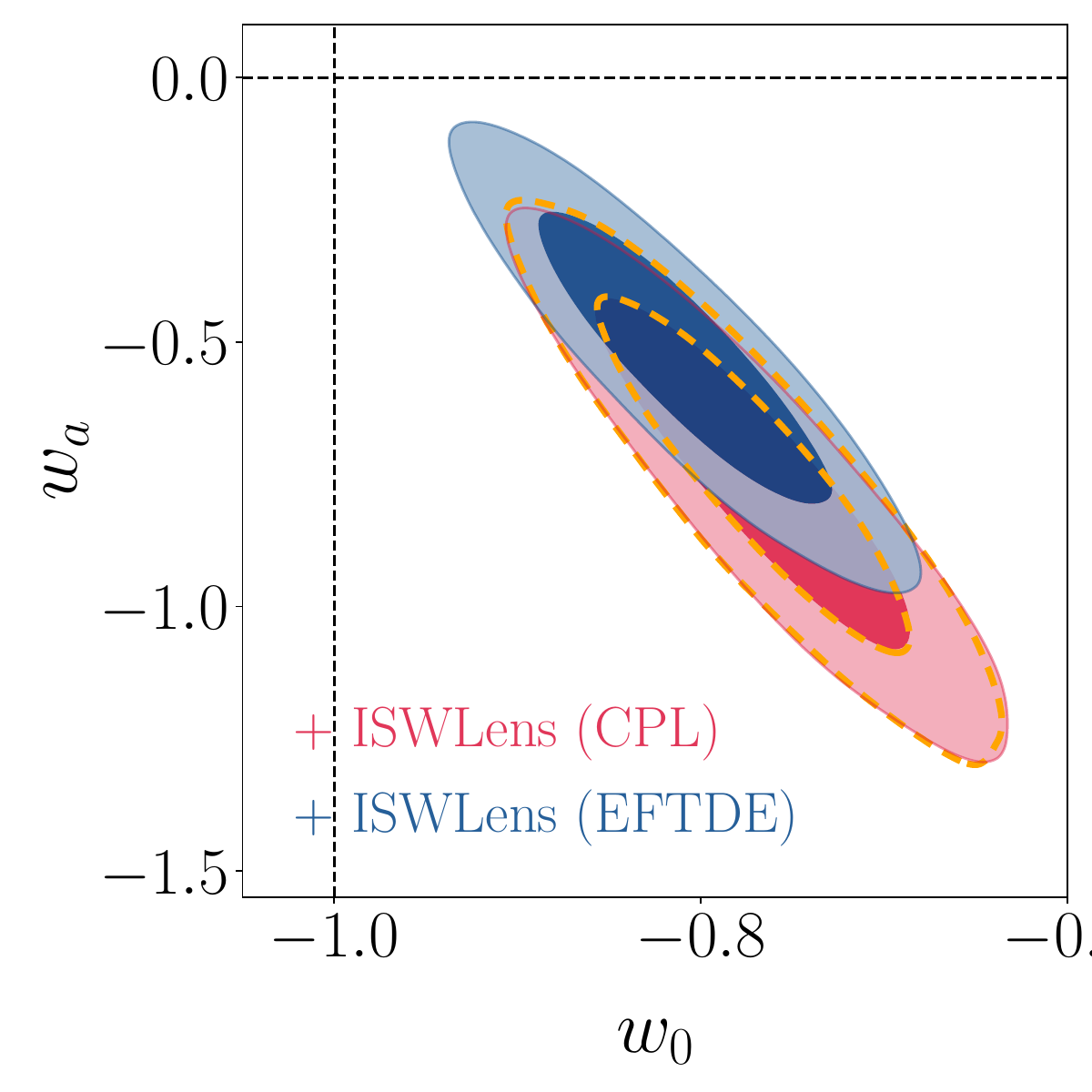}
    \includegraphics[width=0.24\linewidth]{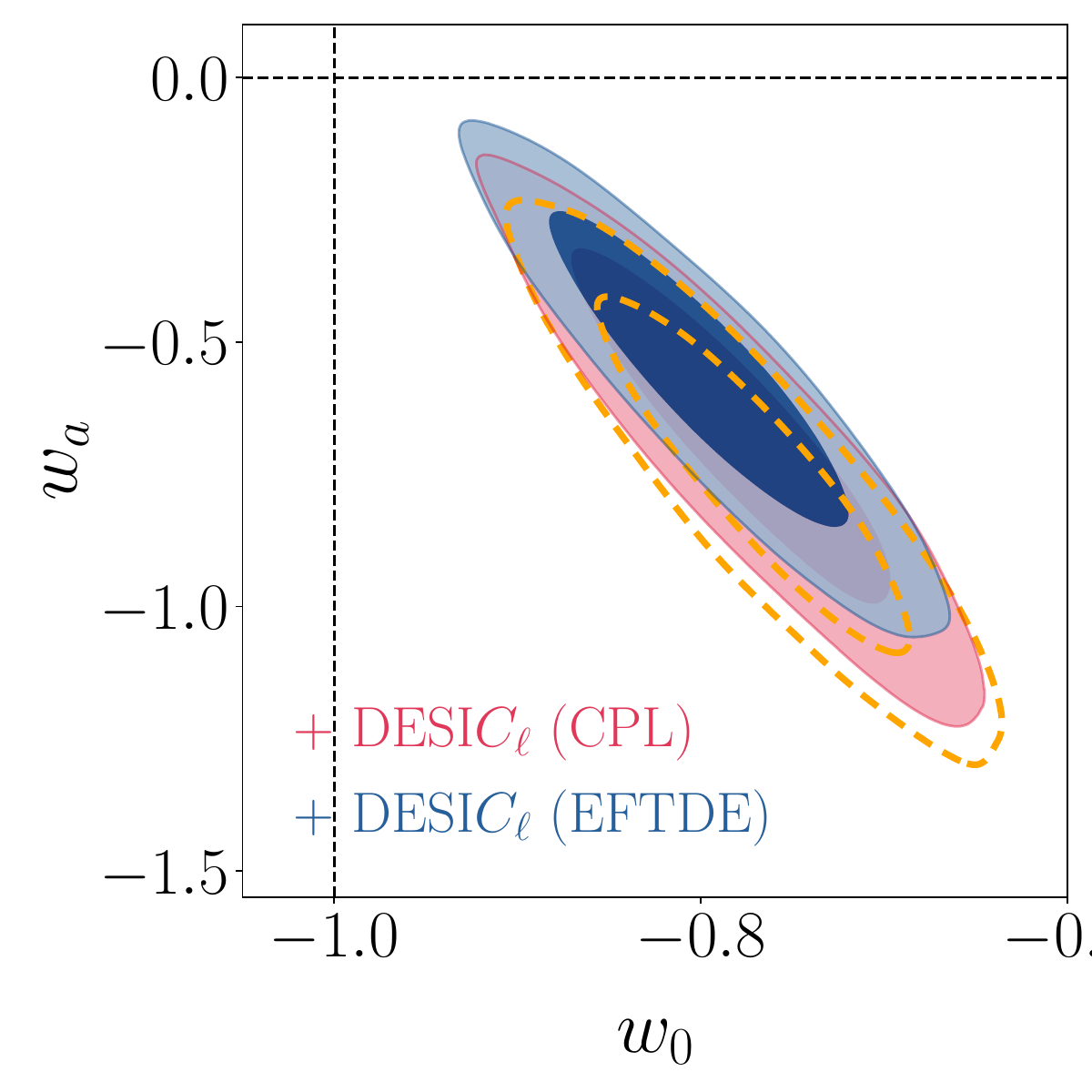}
    \includegraphics[width=0.24\linewidth]{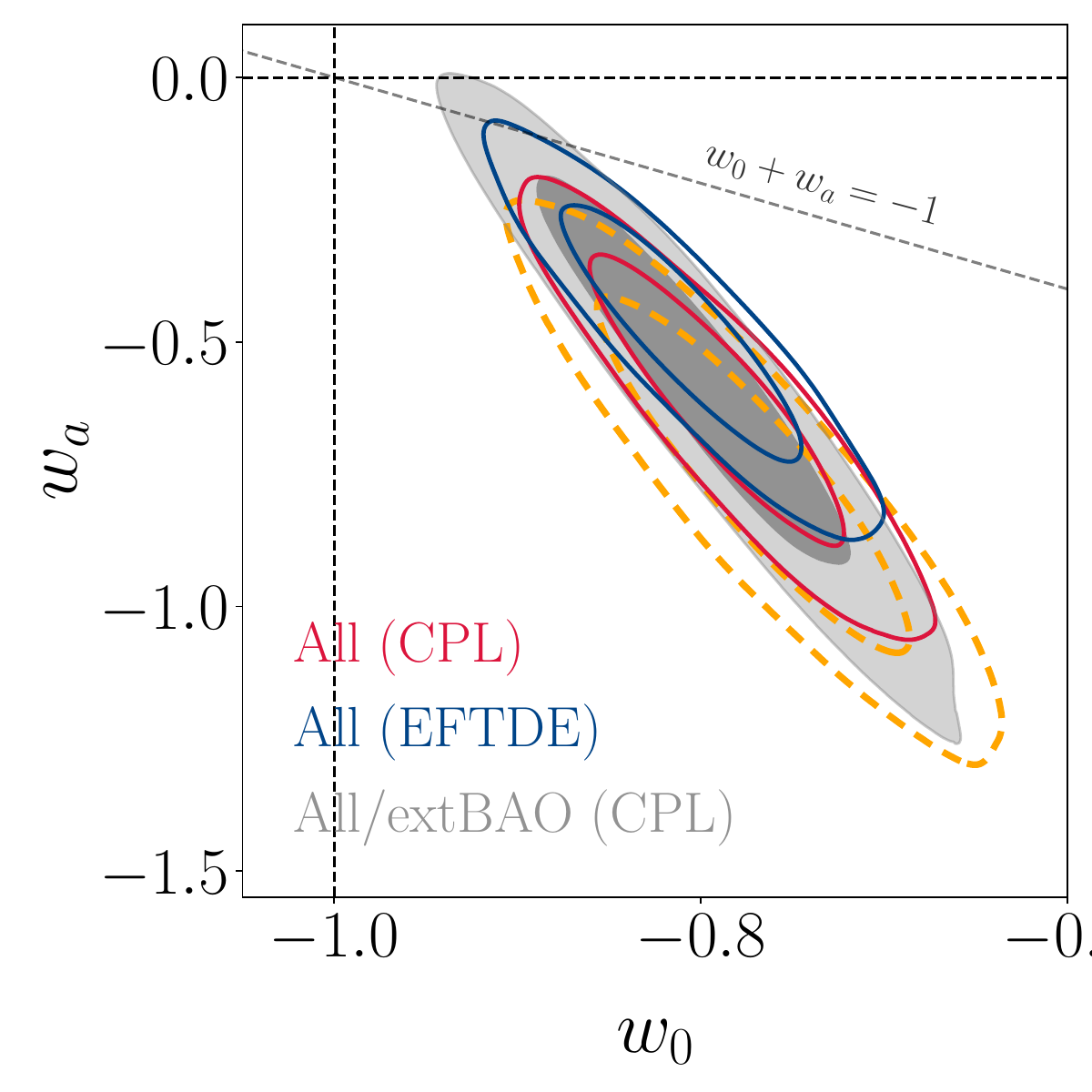}

    % Second row
    \includegraphics[width=0.24\linewidth]{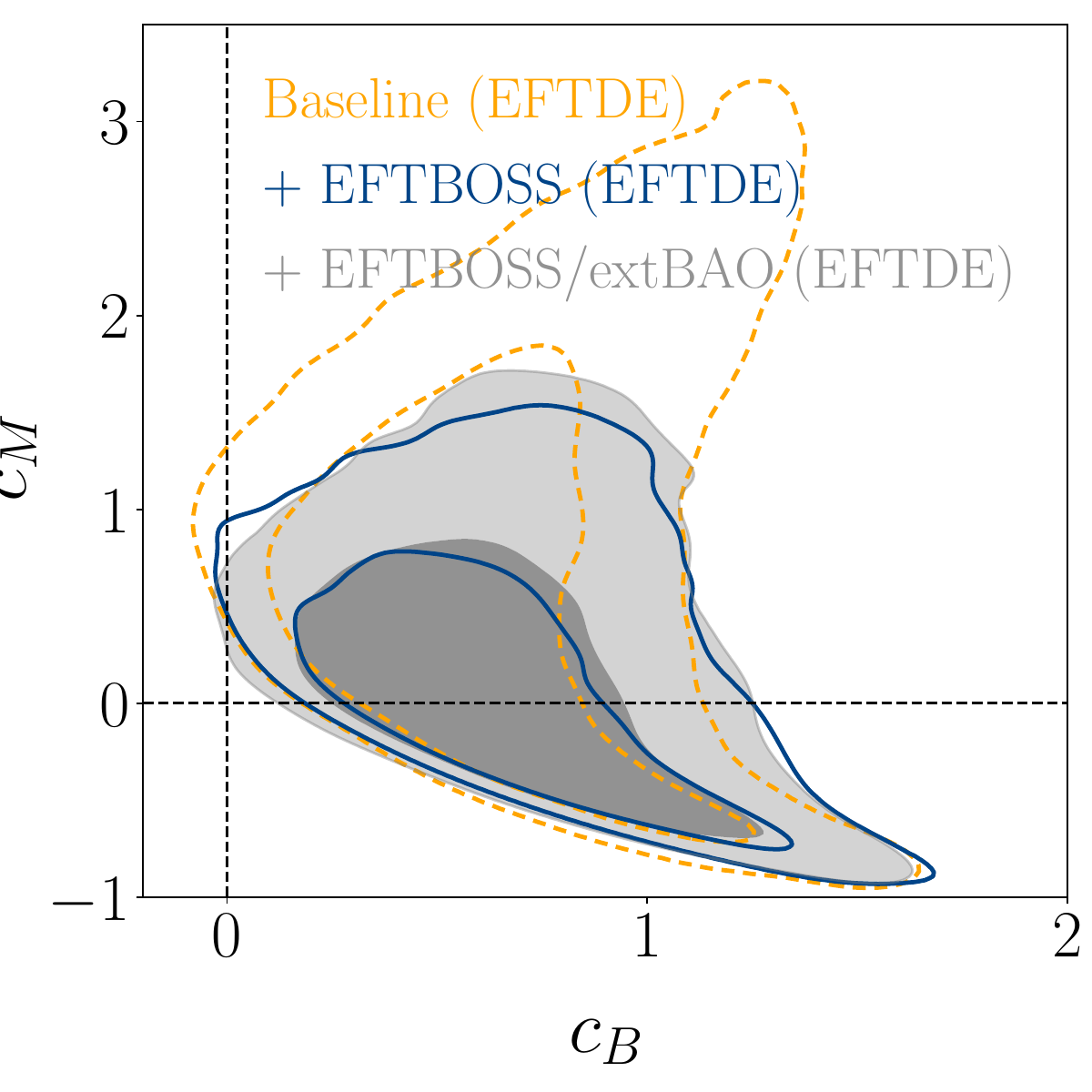}
    \includegraphics[width=0.24\linewidth]{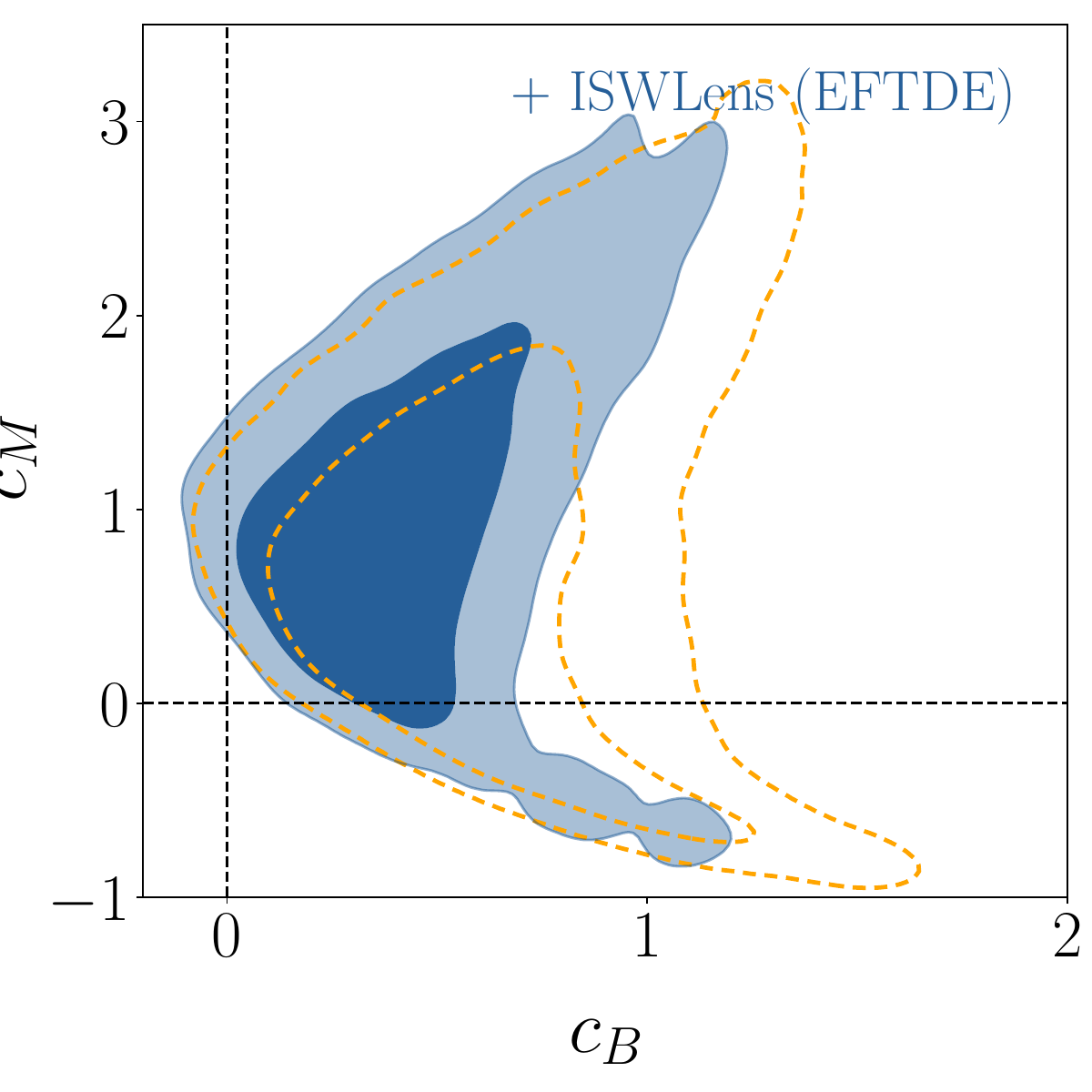}
    \includegraphics[width=0.24\linewidth]{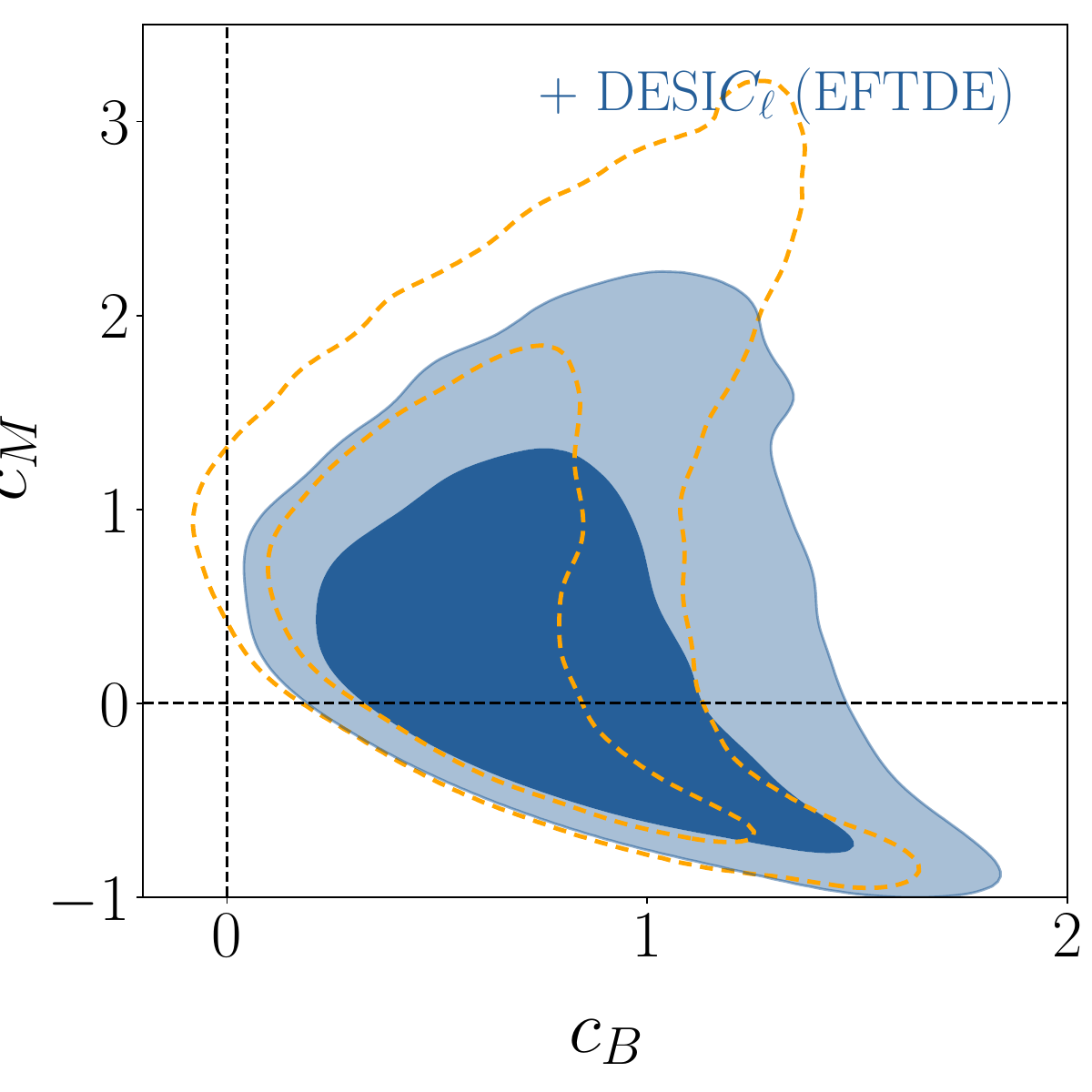}
    \includegraphics[width=0.24\linewidth]{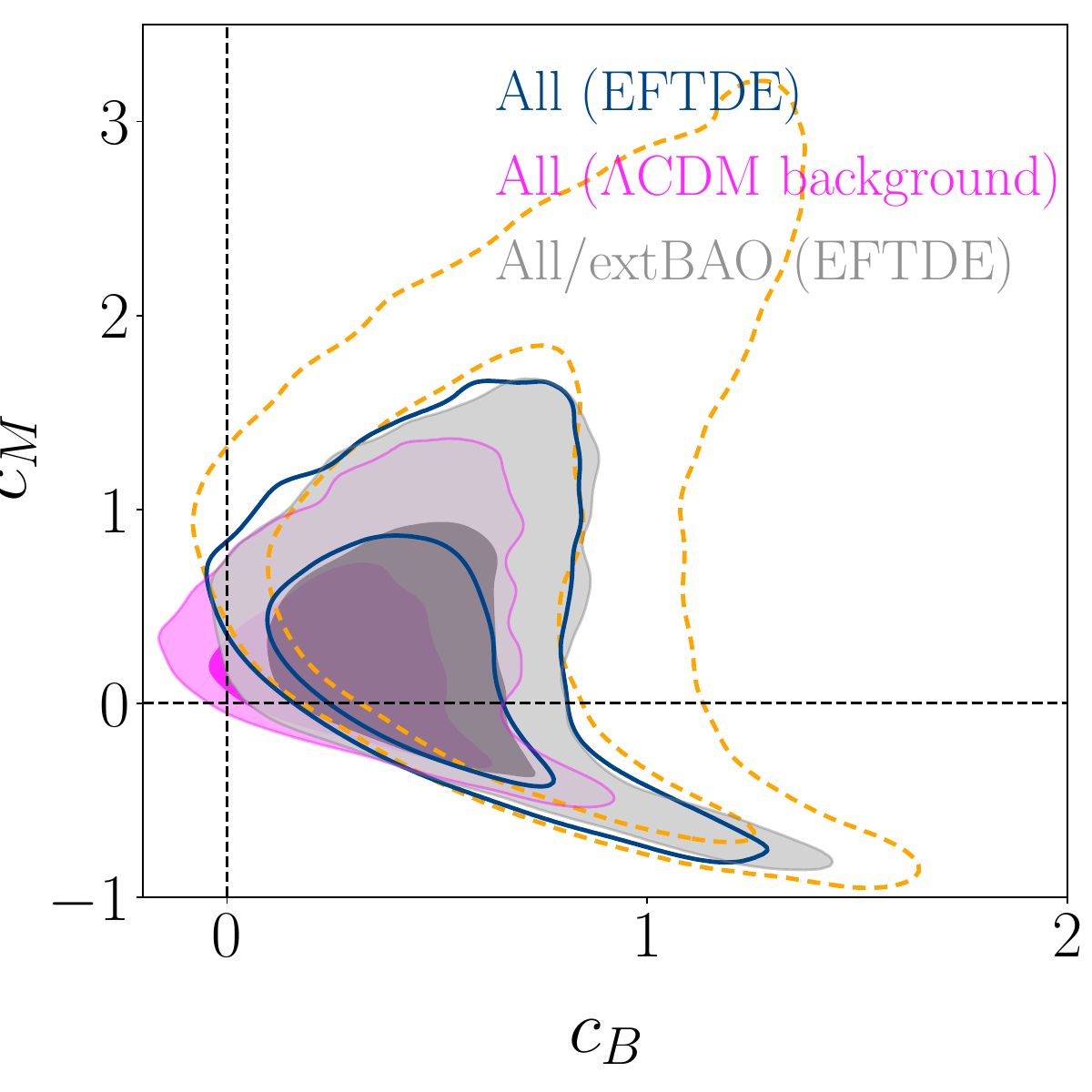}

    \caption{2D posterior distributions of $\{w_0, \, w_a \}$ and $\{c_B, \, c_M \}$ reconstructed from all data combinations considered in this work, namely, Baseline, Baseline + EFTBOSS, Baseline + ISWL, Baseline + DESI$C_\ell$, and the combination of all these datasets. 
    The black dashed lines correspond to the $\Lambda$CDM limits, while the baseline analysis is shown with a yellow dotted line. }
    \label{fig:allcombine}
\end{figure*}

In this section, we discuss the cosmological constraints on the CPL parametrization and on the EFTofDE configuration considered in this work (see Sec.~\ref{sec:theory}).
Our main results are presented in Fig.~\ref{fig:allcombine}, where we show the 2D posterior distributions of $\{w_0, \, w_a \}$ and $\{c_B, \, c_M \}$ for all data combinations presented in Sec.~\ref{sec:analysis_setup}.
The associated cosmological constraints, the best-fit $\chi^2$, and the preferences over $\Lambda$CDM are displayed in Tabs.~\ref{tab:bestfit_params_cpl} and~\ref{tab:bestfit_params_eftde} of App.~\ref{app:supp_results}.
Finally, in Fig.~\ref{fig:baseline_all_w0wa_mg} of that appendix, we also plot the 1D and 2D posterior distributions for all cosmological parameters.

\subsection{Baseline analysis and CMB likelihoods}

\begin{figure}
    \centering
    \includegraphics[width=1\linewidth]{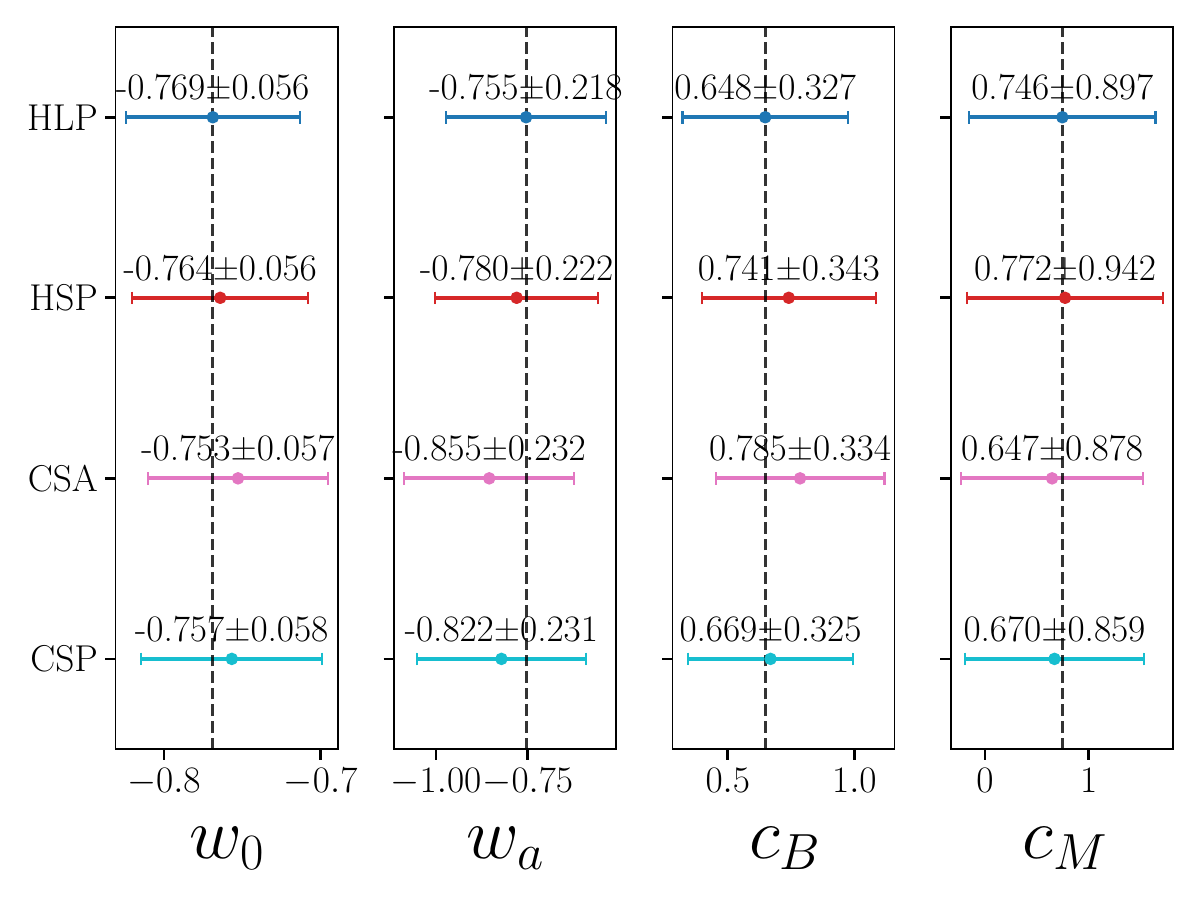}
    \caption{Constraints on $\{w_0, \, w_a, \, c_B, \, c_M \}$ for several CMB likelihood combinations, namely, Hillipop (H) vs Camspec (C) for the high-$\ell$ TTTEEE likelihood, Lollipop (L) vs Simall (S) for the low-$\ell$ TT likelihood, and \textit{Planck} PR4 (P) vs ACT DR6 (A) for the lensing likelihood.
    Note that, for all the data combinations, we use the low-$\ell$ TT likelihood from Commander, together with DESI DR2 BAO and DES Y5. In the first two columns, we assume the CPL parametrization without EFTofDE, while in the last two columns we consider our EFTofDE configuration (with a CPL background). The black dashed lines represent the mean values of our baseline analysis.}
    \label{fig:compare_cmb}
\end{figure}

For the baseline dataset, corresponding to \textit{Planck} PR4 + Lensing + DESI DR2 BAO + DES Y5, we obtain $w_0=-0.769\pm 0.056$ and $w_a=-0.75\pm 0.22$ at $68\%$ CL (with PPF), and a $3.8 \sigma$  preference over $\Lambda$ (see App.~\ref{app:supp_results}).
When we extend this analysis to an EFT description of the DE perturbations, the constraint on $\{w_0, \, w_a \}$ is shifted by $1.2\sigma$ toward its $\Lambda$CDM limit.
For the EFTofDE parameters, we obtain $c_B =  0.65^{+0.24}_{-0.37}$ and $c_M =  0.75^{+0.63}_{-1.0}$, leading to a $3.4 \sigma$ preference over $\Lambda$CDM (see App.~\ref{app:supp_results}). 
While $c_M$ is fully consistent with 0, we observe that $c_B$ is shifted by $1.7\sigma$ compared to its GR limit, which is related to the lensing anomaly~\cite{Tristram:2023haj,DiValentino:2015bja,Calabrese:2008rt}, in line with previous results~\cite{Noller:2018wyv,Specogna:2024euz,Ishak:2024jhs}.\footnote{Although this anomaly has been reduced from $2.8\sigma$ in PR3 to $1.7\sigma$ in \texttt{Camspec}~\cite{Tristram:2020wbi, Tristram:2023haj} and to $0.7\sigma$ in \texttt{Hillipop}~\cite{Rosenberg:2022sdy}.}
Finally, we note that the correlation between $\{ w_0, \, w_a \}$ and $\{ c_B, \, c_M \}$ is weak, as shown in Fig.~\ref{fig:baseline_all_w0wa_mg}.

Compared with the equivalent CPL analysis (without EFTofDE) from the DESI DR2 cosmological results~\cite{DESI:2025zgx}, the best-fit value of $w_a$ shows a mild shift {toward $\Lambda$CDM}, while the evidence for evolving dark energy decreases from $4.2\sigma$ to $3.8\sigma$. This difference arises from the choice of the CMB likelihoods, for both the lensing and the primary spectra.
In Fig.~\ref{fig:compare_cmb}, we compare some variations of our baseline analysis with different CMB likelihoods, namely, Hillipop~\cite{Tristram:2023haj} vs Camspec~\cite{Rosenberg:2022sdy} for the high-$\ell$ TTTEEE likelihood, Lollipop~\cite{Tristram:2023haj} vs Simall~\cite{Planck:2019nip} for the low-$\ell$ TT likelihood, and \textit{Planck} PR4~\cite{Carron:2022eyg} vs ACT DR6~\cite{ACT:2023dou,ACT:2023kun} for the lensing likelihood. We summarize our results in the following:
\begin{itemize}
    \item \textbf{Hillipop vs Camspec:} We observe a shift up to $0.2 \sigma$ on the 1D posteriors of $\{ w_0, \, w_a, \, c_B, \, c_M \}$ (see HSP vs CSP), and a change in the error bars up to $10\%$ (for $c_M$).
    \item \textbf{Lollipop vs Simall:} Lollipop gives slightly tighter constraints than Simall and shifts the posterior up to $0.3\sigma$ for $c_B$ (see HLP vs HSP).
    \item \textbf{Planck PR4 vs ACT DR6 lensing:} The constraints are similar between the two likelihoods with a shift up to $0.35 \sigma$ for $c_B$ (see CSA vs CSP).
\end{itemize}
In the end, we manage to reproduce the results of Ref.~\cite{DESI:2025zgx}, and the main difference between our baseline analysis (HLP) and this reference (CSA) is a shift in $w_a$ of $0.44\sigma$.
Note that in our baseline analysis, the constraining power on $\{c_B, \, c_M \}$ mostly comes from the late integrated Sachs-Wolfe effect and the CMB lensing~\cite{Planck:2015bue}.
In the following, we perform a multiprobe analysis to improve the constraints on these parameters.

\subsection{Multiprobe analysis}\label{sec:different_datacombination}

\begin{figure*}
    \centering
        \begin{minipage}{1\textwidth}
        \centering
         \includegraphics[width=0.48\textwidth]{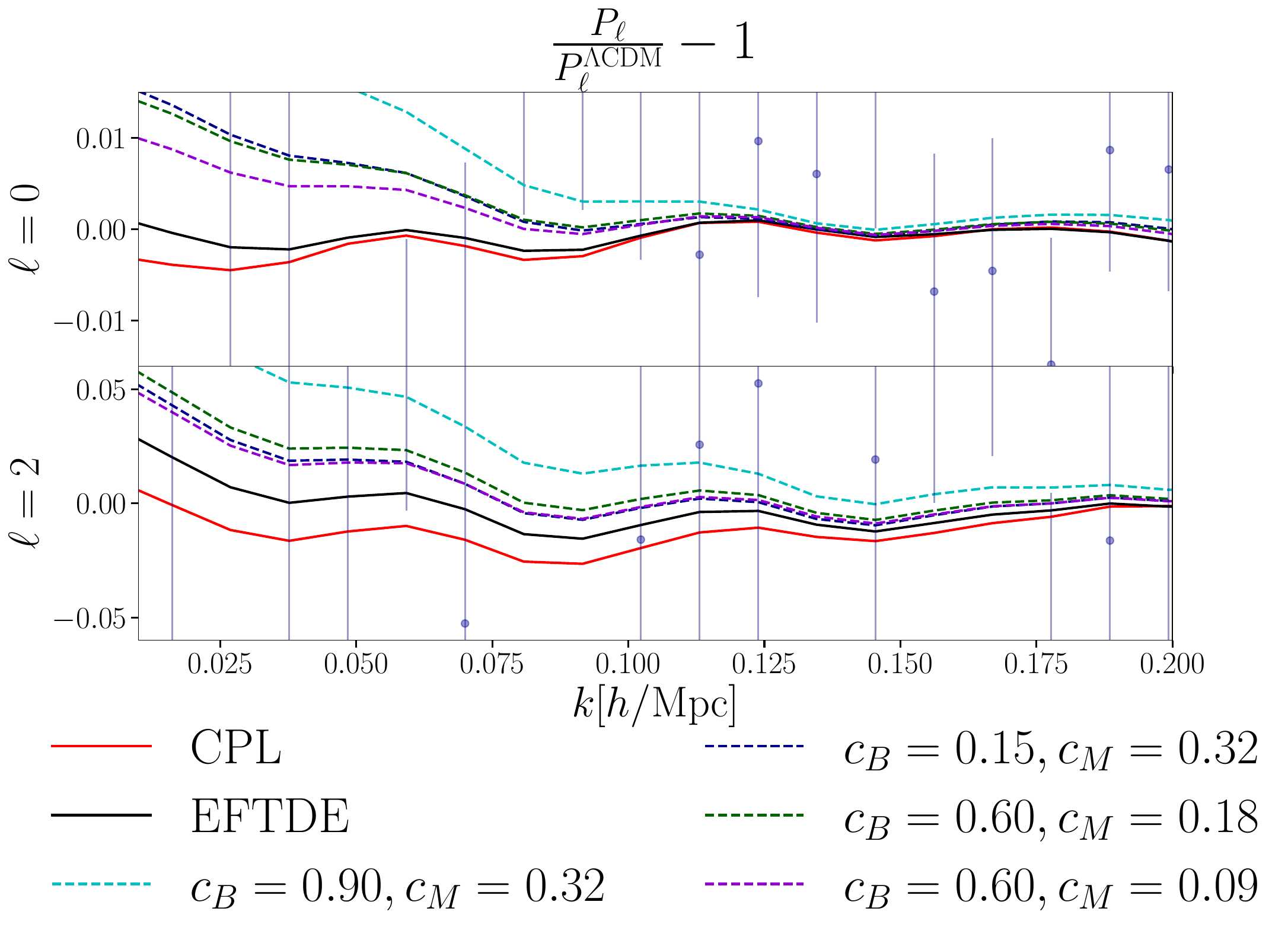}
    \includegraphics[width=0.48\textwidth]{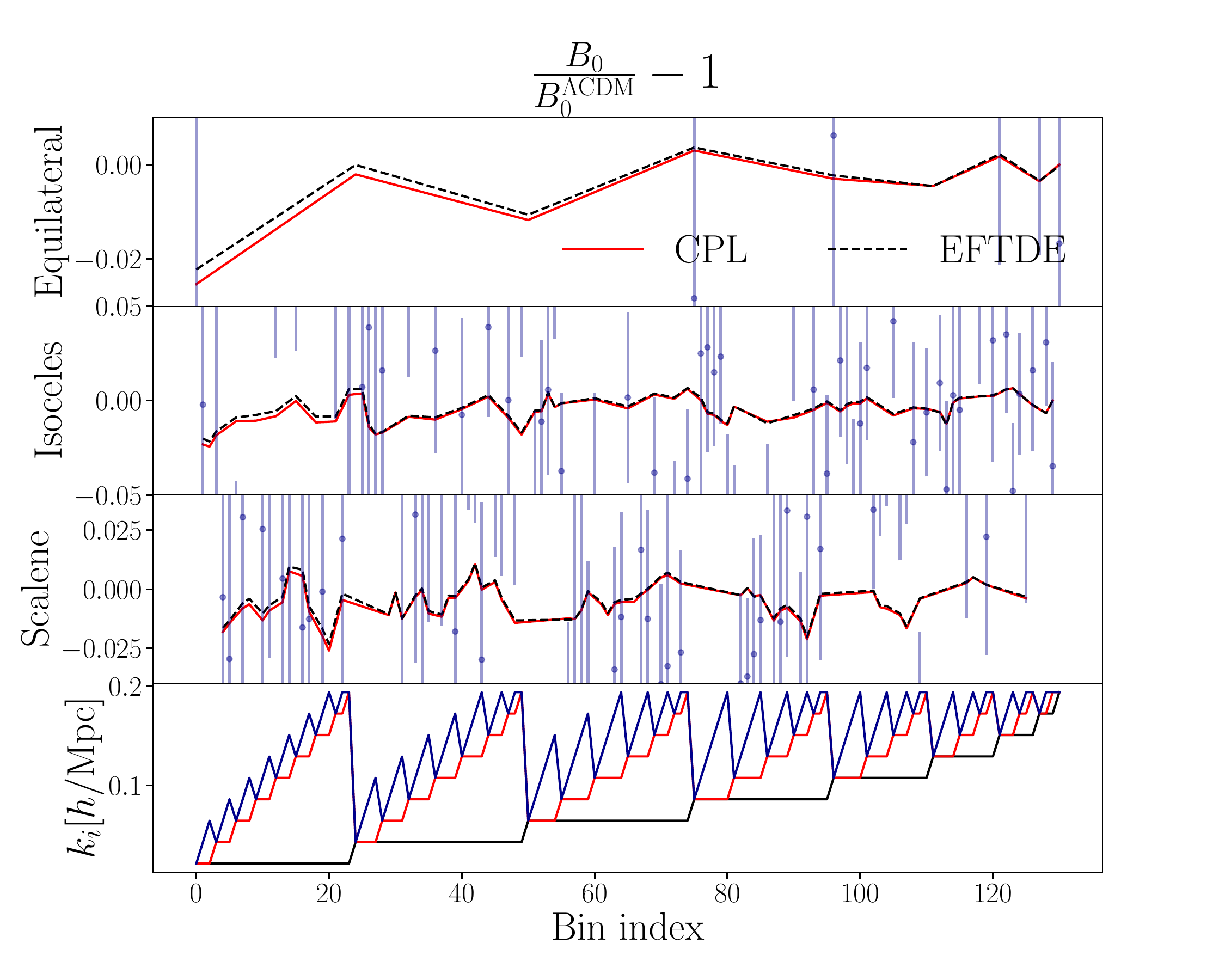}
    \end{minipage}
    \caption{ \textit{Left:} Residuals of the monopole and quadrupole of the galaxy power spectrum for the CPL model with and without EFTofDE, normalized to the $\Lambda$CDM model (for the baseline + EFTBOSS analysis). 
    {We also show the impact of varying $c_B$ and $c_M$ on this observable, setting the other cosmological parameters to the best-fit.}
    Note that we show here the predictions and the data for the low-$z$ NGC sample of BOSS.
    \textit{Right:} Same for the bispectrum monopole for different triangle configurations.}
    
    \label{fig:residual_EFTBOSS}
\end{figure*}

\subsubsection{EFTBOSS} 
We first extend the baseline dataset by incorporating both the power spectrum and bispectrum measurements from BOSS thanks to an EFTofLSS description of the dark matter overdensity field (see Sec.~\ref{sec:analysis_setup}).
Unlike Ref.~\cite{Lu:2025gki}, which employed the exact time dependence description of the EFTofLSS operators, we adopt here the EdS approximation where the time evolution of all the operators is only encoded in the linear growth factor $D_+(a)$, since, as demonstrated in Ref.~\cite{Lu:2025gki}, the difference between the exact-time and EdS treatments is negligible for BOSS sensitivity. We further restrict our analysis to the smooth quintessence case, as the smooth and clustering scenarios~\cite{Sefusatti:2011cm,Creminelli:2009mu} studied in Ref.~\cite{Lu:2025gki} exhibit similar deviations.
Our results are presented in the first column of Fig.~\ref{fig:allcombine}.

The sensitivity of EFTBOSS to $c_B$ and $c_M$ in the quasi-static limit arises from the fact that they act as additional sources in the Poisson equation \cite{Cusin:2017mzw}, as can be seen in Eq.~\eqref{eq:poisson}. 
This implies a modification of the growth factor $D_+(a)$ [and, therefore, of the growth rate $f_+(a)$] through the continuity and Euler equations, which will modify the time dependence of the EFTofLSS operators~\cite{Zhang:2021uyp,Perko:2016puo,Rodriguez-Meza:2023rga}.
These parameters also affect the overall amplitude of the linear power spectrum within the scales of interest, namely, $0.01 < k/[{\rm h/Mpc}] < 0.23$, which propagate into the one-loop corrections. 
In the top panel of 
Fig.~\ref{fig:theory_varying_cBcM}, we display the ratios of the linear matter power spectrum for several values of $c_B$ and $c_M$, showing that variations in these parameters lead to an almost scale-independent modification of the linear matter power spectrum amplitude.
We further note that EFTBOSS is more affected by a change in $c_M$ than a change in $c_B$, coming from the fact that $\mu$ is more sensitive to a variation in the former parameter at high redshift, while a variation in the latter parameter becomes significant only at low redshift (see Fig.~\ref{fig:vary_mu_sigma_against_cBcM}).\footnote{Let us note that since the growth factor is coupled to $\mu$ through a differential equation, the effect of $c_M$ on the matter density perturbations accumulates over cosmic time, implying that EFTBOSS is more sensitive to this parameter than $c_B$.}
{In the left panel of Fig.~\ref{fig:residual_EFTBOSS}, we display the residuals of the monopole and quadrupole of the galaxy power spectrum for several values of $c_B$ and $c_M$, allowing us to stress that they affect the galaxy power spectrum in the same way, namely, that an increase in these parameters involves an enhancement of this observable. This suggests that there is a negative correlation between these two parameters in the EFTBOSS likelihood, and that a slight increase in $c_M$ can be compensated for by a larger decrease in $c_B$.}

When we add EFTBOSS to the baseline analysis (using the $r_s$-marginalization procedure described in Sec.~\ref{sec:analysis_setup}), the constraint on the 2D plane $\{ w_0, \, w_a \}$ is improved by $27\%$,\footnote{Throughout the paper, in order to quantify an improvement in the $\{w_0, \, w_a \}$ or $\{c_B, \, c_M  \}$ 2D planes, we use the ratio of the figure of merit (FOM)~\cite{Albrecht:2006um,DESI:2024hhd}, defined as FOM$\propto |{\rm det} C|^{-1/2}$, where $C$ is the covariance matrix of the parameter posteriors. {This metric should be taken with a grain of salt for the $\{c_B, \, c_M \}$ 2D posterior distributions, given that they are not Gaussian.}} {and the preference over $\Lambda$ increases from $3.7 \sigma$ to $4.7 \sigma$}, consistent with previous findings~\cite{Lu:2025gki}. Within the EFTofDE framework, we obtain a $10.3\%$ ($84.7\%$) improvement for the $\{w_0, \, w_a \}$ ($\{c_B, \, c_M \}$) plane, increasing the preference over $\Lambda$CDM from $3.4 \sigma$ to $4.3 \sigma$. 
As shown in Fig.~\ref{fig:baseline_all_w0wa_mg}, EFTBOSS is able to break the degeneracy between $\sigma_8$ and $c_M$, leading to a $38 \%$ improvement in the constraint of $c_M$ (compared to $2 \%$ for $c_B$), allowing us to exclude large positive values of the running of the Plank mass, namely, $c_M \gtrsim 1.5$ (see Fig.~\ref{fig:allcombine}).
{Finally, and as expected, EFTBOSS induces a negative correlation between $c_B$ and $c_M$ (see Fig.~\ref{fig:allcombine}) due to the exclusion of high values of $c_M$ (for which the galaxy power spectrum is more sensitive).}

To further investigate the constraining power of EFTBOSS, we show, in Fig.~\ref{fig:residual_EFTBOSS}, the power spectrum and bispectrum residuals for the best-fit of CPL with and without EFTofDE, normalized to the equivalent $\Lambda$CDM best-fit.
For the CPL model with PPF, we obtain $\delta \chi_{\rm min}^2{\rm (EFTBOSS)} = - 7.4$ with respect to $\Lambda$CDM (within the baseline + EFTBOSS analysis), corresponding to a non-negligible  improvement of the fit to the BOSS data.
Within the EFTofDE framework, we do not further improve this fit (as can be seen in Fig.~\ref{fig:residual_EFTBOSS}), with $\delta \chi_{\rm min}^2 = - 7.5$.

Using the extBAO dataset (instead of DESI BAO DR2), and without considering the $r_s$-marginalization procedure for the EFTBOSS likelihood,  we observe a very similar trend, as shown in Fig.~\ref{fig:allcombine}.
The extBAO likelihood provides weaker constraints on the late-time expansion than the DESI DR2 BAO likelihood, leading to a smaller constraint on the $\{ w_0, \,w_a \}$ plane, although consistent with previous analyses.
However, the constraints on $c_B$ and $c_M$ remain comparable, resulting from their weak correlation with the background parameters, especially $\Omega_m$, $w_0$ and $w_a$ (see Fig.~\ref{fig:baseline_all_w0wa_mg}).
We note that the statistical significance for evolving dark energy decreases from $4.7\sigma$ to $3.6\sigma$ with PPF and from $4.3\sigma$ to $3.3\sigma$ with EFTofDE.

\subsubsection{ISWL}

\begin{figure}
    % \centering
    \begin{minipage}{0.5\textwidth}
        \centering
         \includegraphics[width=1.\linewidth]{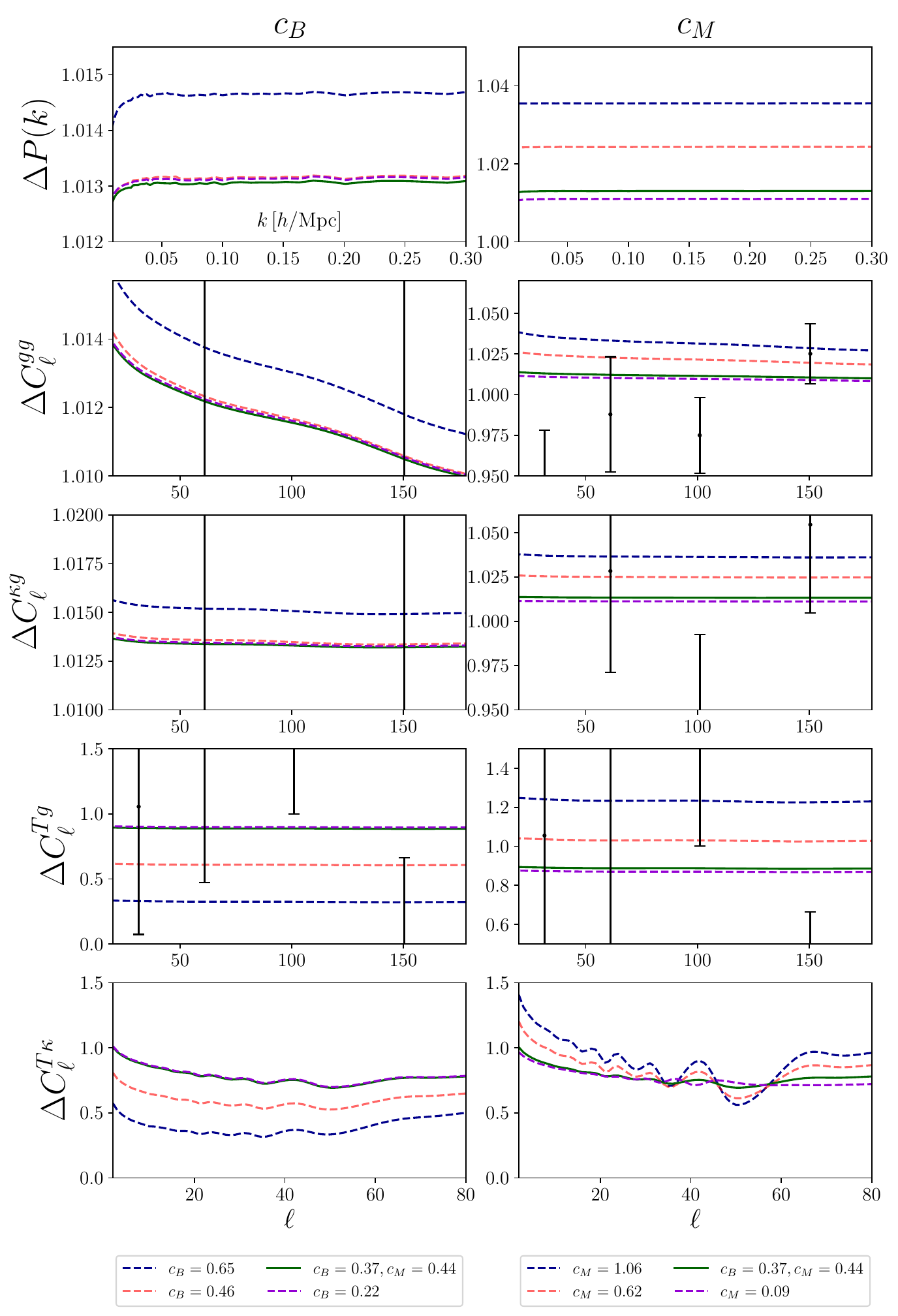}
    \end{minipage}
    \caption{Residuals (with respect to $\Lambda$CDM) of the linear matter power spectrum, $C_\ell^{gg}$, $C_\ell^{\kappa g}$, $C_\ell^{Tg}$, and $C_\ell^{T \kappa}$ for several values of $c_B$ (\textit{left}) and $c_M$ (\textit{right}).
    The residuals are computed at $z=0.625$, corresponding to the effective redshift of the second redshift range of the DESI$C_\ell$ likelihood.
    All cosmological parameters are fixed to their best-fit values (reconstructed from the ``All'' analysis), while the solid green curves correspond to the best-fit values of $c_B$ and $c_M$.
    For the sake of clarity, we divide the error bars of $C_\ell^{\kappa g}$ and $C_\ell^{Tg}$ by $2$ (see Fig.~\ref{fig:namaster_cltg} for a real representation of these data points).}
    \label{fig:theory_varying_cBcM}
\end{figure}

In Fig.~\ref{fig:theory_varying_cBcM}, we show the ratios of $C_\ell^{T\kappa}$ for several values of $c_B$ and $c_M$, allowing us to highlight that this observable is highly sensitive to these parameters.
In particular, the more we increase the value of the braiding parameter $c_B$, the more the $C_\ell^{T\kappa}$ power spectrum is suppressed, which is explained by the fact that this parameter reduces the integrated Sachs-Wolfe effect (and can even make it negative for large positive values of $c_B$)~\cite{Renk:2016olm,Zumalacarregui:2016pph,Seraille:2024beb,Chudaykin:2025gdn}.
For the running of Planck mass $c_M$, we observe the opposite behavior at large scales, implying that the ISWL likelihood introduces a positive correlation between these two EFT parameters (see Fig.~\ref{fig:allcombine}).

When we add the ISWL likelihood to the baseline analysis, we find that the constraints on $c_B$ and $c_M$ are improved by $25\%$ and $16\%$, respectively, while the background dark energy parameters remain essentially unchanged (see Fig.~\ref{fig:allcombine}). 
As expected, we can see in Fig.~\ref{fig:allcombine} that this likelihood induces a positive correlation in the $\{c_B, c_M  \}$ plane, which is significantly reduced by $51.6 \%$.
This positive correlation is accompanied by the exclusion of large positive values of the braiding parameter, namely, $c_B \gtrsim 1$, which is consistent with Ref.~\cite{Chudaykin:2025gdn}. 
In addition, we note that the inclusion of the ISWL likelihood does not impact the preference for evolving dark energy, which remains at $3.8 \sigma$ ($3.4\sigma$) for the analysis without (with) the EFT description of the DE perturbations.

\subsubsection{DESI$C_\ell$}

In Fig.~\ref{fig:theory_varying_cBcM}, we also show the ratios of $C_\ell^{gg}$, $C_\ell^{\kappa g}$, and $C_\ell^{T g}$ for several values of $c_B$ and $c_M$. Interestingly, a variation of these parameters induces a scale-independent modification of the amplitude of these three observables, with different correlations between $c_B$ and $c_M$.
In particular, $C_\ell^{Tg}$ exhibits the same behavior as $C_\ell^{T\kappa}$ (see Ref.~\cite{Seraille:2024beb}) since both are directly sensitive to $\Sigma$, as shown in Eq.~\eqref{eq:cltg_kernel}, therefore introducing a positive correlation between the two EFT parameters.
However, we observe a different trend for $C_\ell^{gg}$ and $C_\ell^{\kappa g}$, where an increase in both $c_B$ or $c_M$ leads to an enhancement of these observables, implying a negative correlation between these parameters.
Interestingly, we expect that combining $C_\ell^{gg}$, $C_\ell^{\kappa g}$, $C_\ell^{T g}$, and $C_\ell^{T \kappa}$ will allow us to break the various degeneracies described above. 

When we add the DESI$C_\ell$ likelihood to the baseline dataset (where we have removed the low-$\ell$ CMB lensing data as explained in Sec.~\ref{sec:data_corr}), the precision on the $\{w_0, \, w_a \}$ plane remains equivalent to the baseline analysis (regardless of the DE perturbation parametrization), though the mean values are slightly shifted toward $\Lambda$CDM: $w_0 = -0.769 \pm 0.056 \rightarrow -0.780 \pm 0.056$ and $w_a = -0.75 \pm 0.22 \rightarrow -0.67 \pm 0.22$ (see Fig.~\ref{fig:allcombine}).
Regarding the EFTofDE parameters, the constraint on
$c_M$ is improved by $17 \%$, while the constraint on $c_B$ remains the same.
Indeed, as shown in Fig.~\ref{fig:allcombine}, the DESI$C_\ell$ likelihood is able to set a stronger constraint on the upper limit of $c_M$ compared to the baseline analysis, excluding the region $c_M \gtrsim 2$, consistent with Ref.~\cite{Seraille:2024beb}.
As a result, we can clearly see a negative correlation between the two EFT parameters, coming from  the fact that the main constraining power comes from $C_\ell^{gg}$.
This is expected given the small error bars of this observable compared to $C_\ell^{Tg}$ and $C_\ell^{\kappa g}$ (see Fig.~\ref{fig:namaster_cltg} of App.~\ref{appendix:mapar}).
Finally, we note that the preference for evolving dark energy remains similar to the baseline analysis.

Our analysis can be improved by including the low-$\ell$ data points of the CMB lensing, after determining the covariance between $C_\ell^{\kappa \kappa}$ and $C_\ell^{\kappa g}$.
We further note that including smaller scales in $C_\ell^{gg}$ would help to improve the constraints on the EFTofDE parameters (see Fig.~\ref{fig:theory_varying_cBcM}).
We leave this exploration for future work, using a nonlinear galaxy bias expansion, as done in Ref.~\cite{Sailer:2024jrx}.

\subsubsection{All}\label{sec:res_all}

In the previous sections, we saw that EFTBOSS and DESI$C_\ell$ introduce a negative correlation between $c_B$ and $c_M$, by imposing a stronger constraint on the upper limit of $c_M$ compared to the baseline analysis (see Fig.~\ref{fig:allcombine}).
Conversely, the ISWL likelihood introduces a positive correlation between these two parameters, by imposing a stronger constraint on the upper limit of $c_B$ compared to the baseline analysis (see Fig.~\ref{fig:allcombine}).
Therefore, we expect that combining all these likelihoods will further break these degeneracies, and then improve the constraints on $c_B$ and $c_M$.

When we perform such an analysis (always by removing the low-$\ell$ CMB lensing data), we improve the constraint on the $\{w_0, \, w_a \}$ plane by $46 \%$ ($52 \%$) for the analysis without (with) EFTofDE compared to the baseline analysis. This is accompanied by a $33 \%$ ($35 \%$) reduction of the $\{ h, \, \Omega_m \}$ 2D posterior distribution.
Regarding the EFTofDE parameters, the $\{c_B, \, c_M \}$ plane is reduced by $177 \%$ compared to the baseline analysis, and we obtain $c_B = 0.46^{+0.16}_{-0.22}$ and $c_M = 0.31^{+0.39}_{-0.49}$, respectively corresponding to an improvement of $37 \%$ and $46 \%$.
As explained above, this strong constraint stems from the different degeneracies between the EFTofDE parameters that are inherent in the different likelihoods considered in our work.
We finally obtain a preference for evolving dark energy of $4.6 \sigma$ ($4.2 \sigma$) for the analysis with PPF (EFTofDE), compared to $3.8 \sigma$ ($3.4 \sigma$) for the baseline analysis, while we note that the 2D posterior distribution of $\{c_B, \, c_M  \}$ is compatible with its GR limit at $\sim 2 \sigma$.
{We note that the posterior of $\{ w_0, w_a \}$ lies at $3.5\sigma$ ($3.1\sigma$) from the $w_0 + w_a = -1$ line, corresponding to a transition from $w < -1$ in the past to $w > -1$ today (see quintom models~\cite{Cai:2025mas,Cai:2009zp,Feng:2004ad,Xia:2007km,Zhao:2012aw,Guo:2004fq,Zhao:2005vj,Xia:2005ge}).}
To understand the origin of this preference, we display in Tab.~\ref{tab:bestfit_params_eftde_transposed} the best-fit $\chi^2$ for each likelihood for $\Lambda$CDM, CPL with PPF, and CPL with EFTofDE. We obtain a total $\Delta \chi^2_{\rm min} = - 24.6$ ($\Delta \chi^2_{\rm min} = - 26.2$) for the CPL model with PPF (EFTofDE), mostly coming from the EFTBOSS ($\delta \chi^2_{\rm min} \sim -7 $), DESI BAO ($\delta \chi^2_{\rm min} \sim - 5$), and DES Y5 ($\delta \chi^2_{\rm min} \sim - 12 $) likelihoods.

To investigate the dependence of the $\{c_B, \, c_M  \}$ 2D posterior distribution on the background cosmology, we perform the same analysis fixing the DE background evolution to $\Lambda$, as shown in Fig.~\ref{fig:allcombine}.
We obtain $c_B = 0.32^{+0.18}_{-0.21}$ and $c_M = 0.25^{+0.25}_{-0.43}$, respectively corresponding to a better constraint of $10 \%$ and  $22 \%$ compared to the analysis with the CPL background evolution.
We further note that the $\{c_B, \, c_M  \}$ plane is slightly shifted toward its GR limit, corresponding to a preference over $\Lambda$CDM of $0.68 \sigma$ (and a $\chi^2$ improvement of $-1.4$).

\subsection{Impact of BAO and SN data}\label{sec:res_nobaonosn}

\begin{figure}
    \centering
    \begin{minipage}{0.5\textwidth}
        \includegraphics[width=0.45\linewidth]{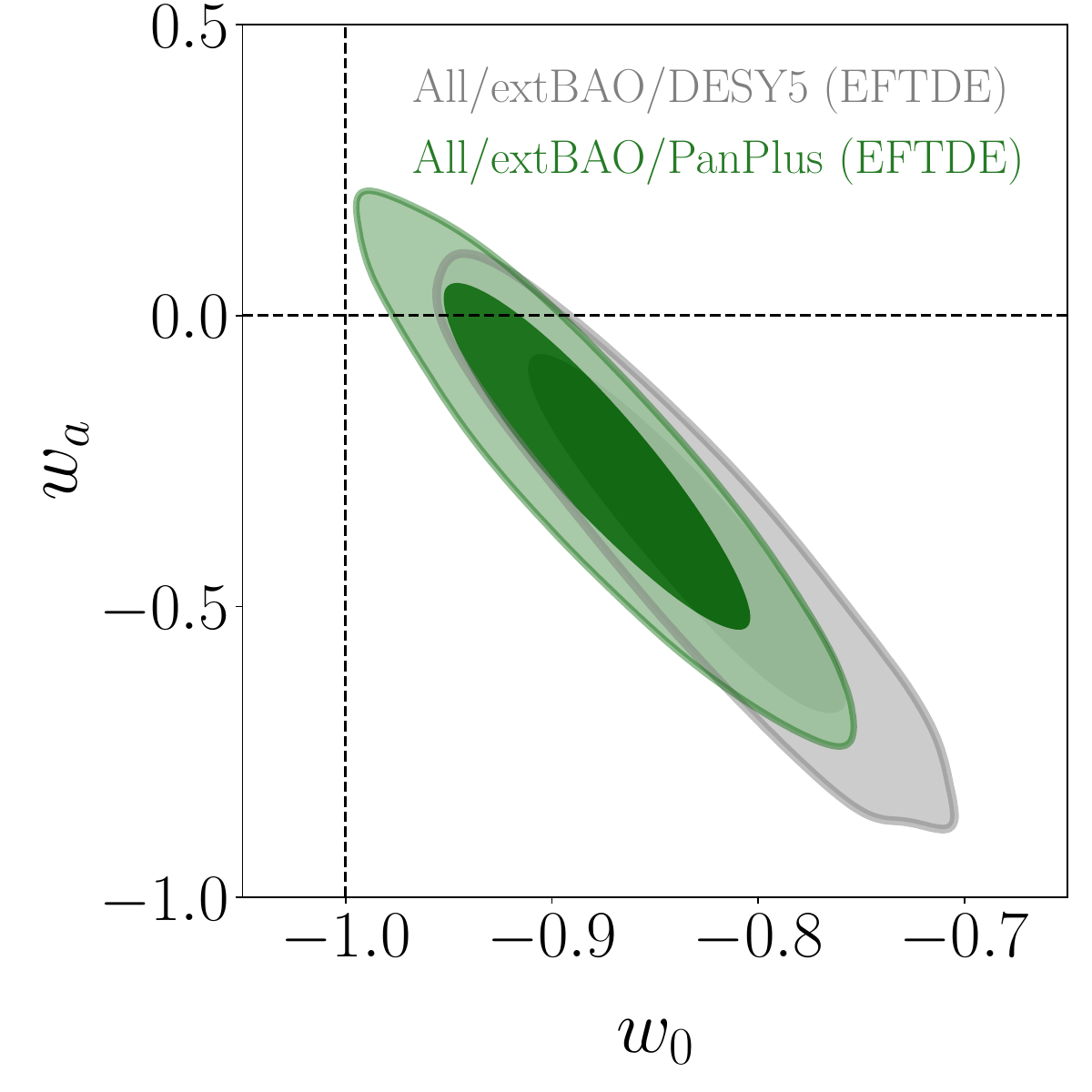}
    \includegraphics[width=0.45\linewidth]{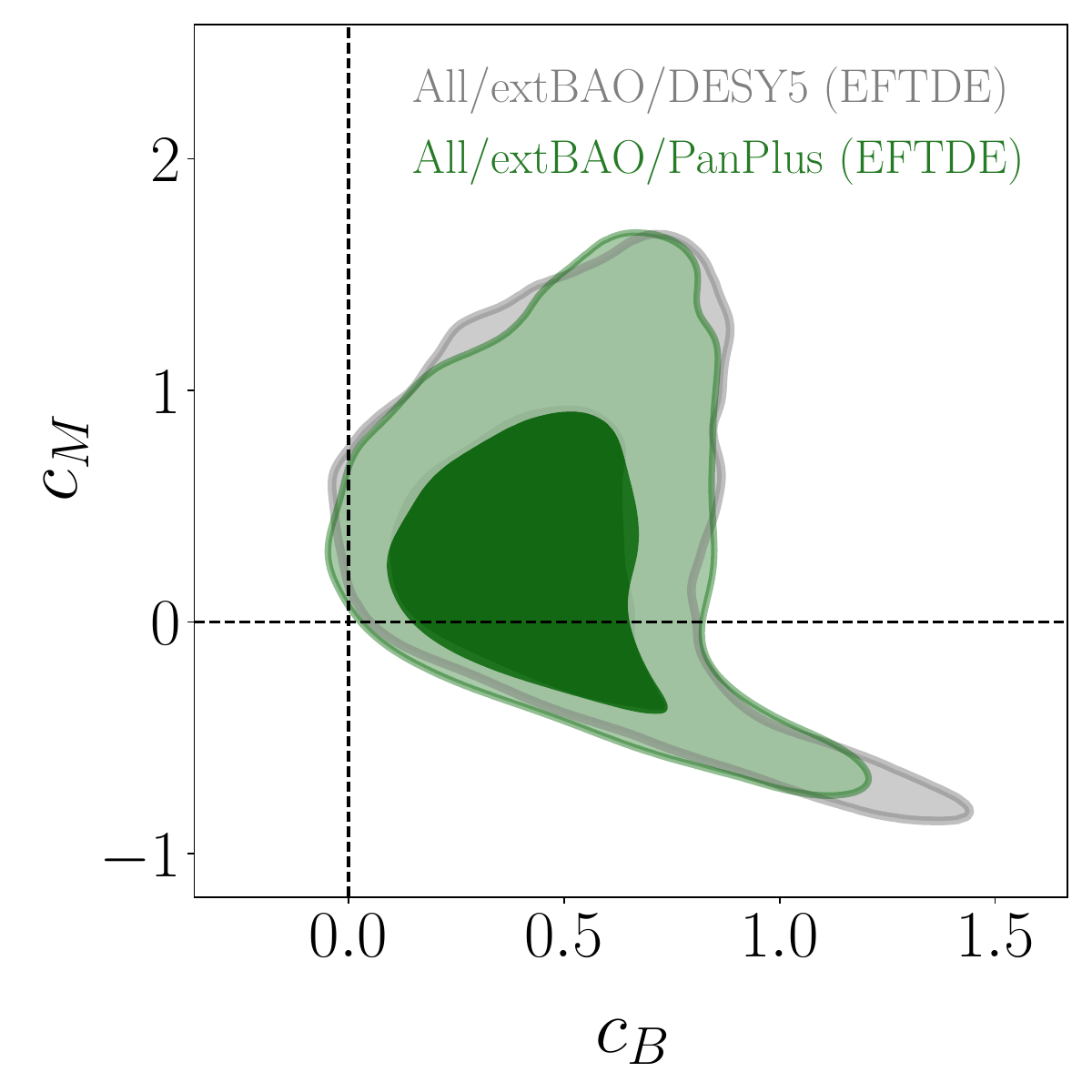}
    \includegraphics[width=0.45\linewidth]{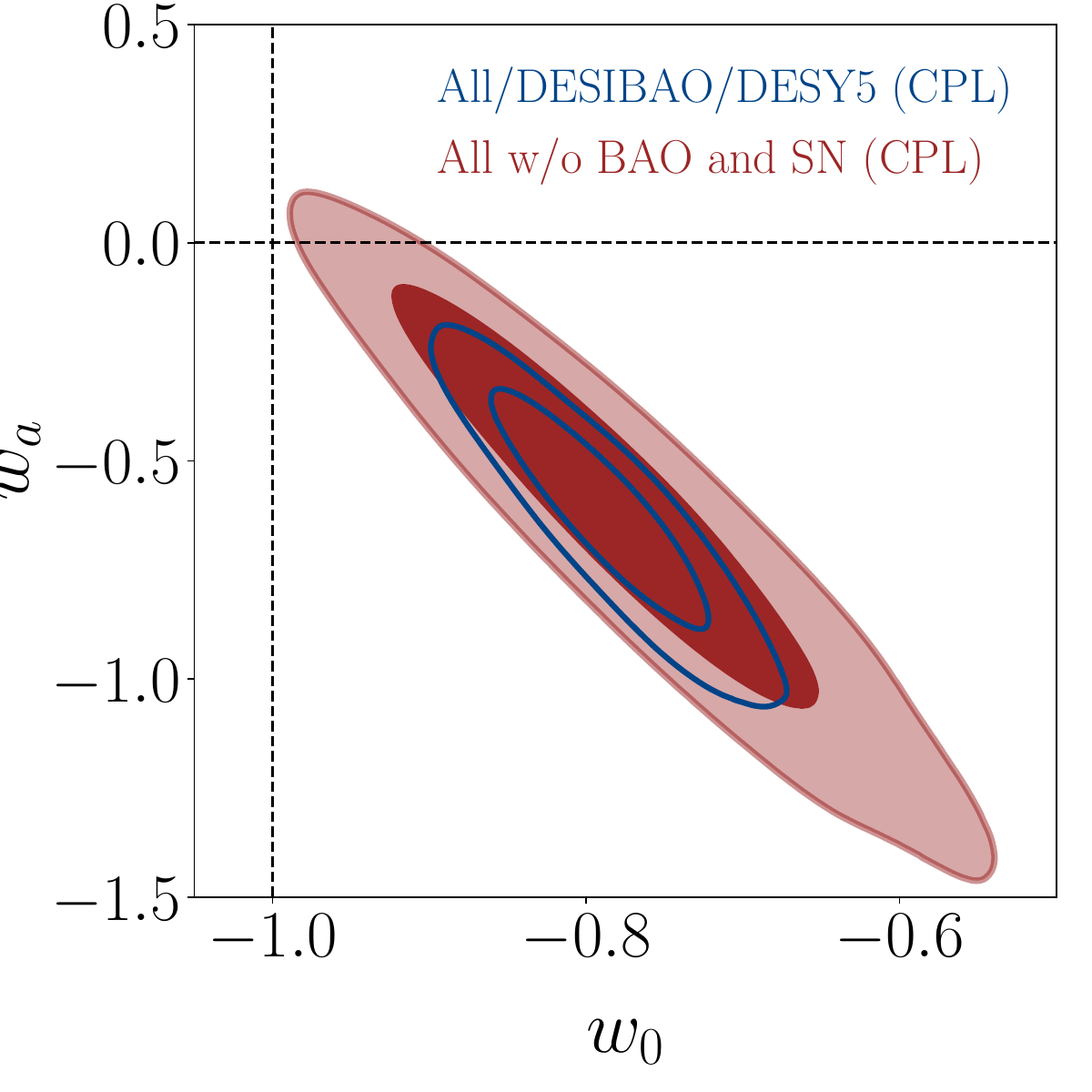}
    \includegraphics[width=0.45\linewidth]{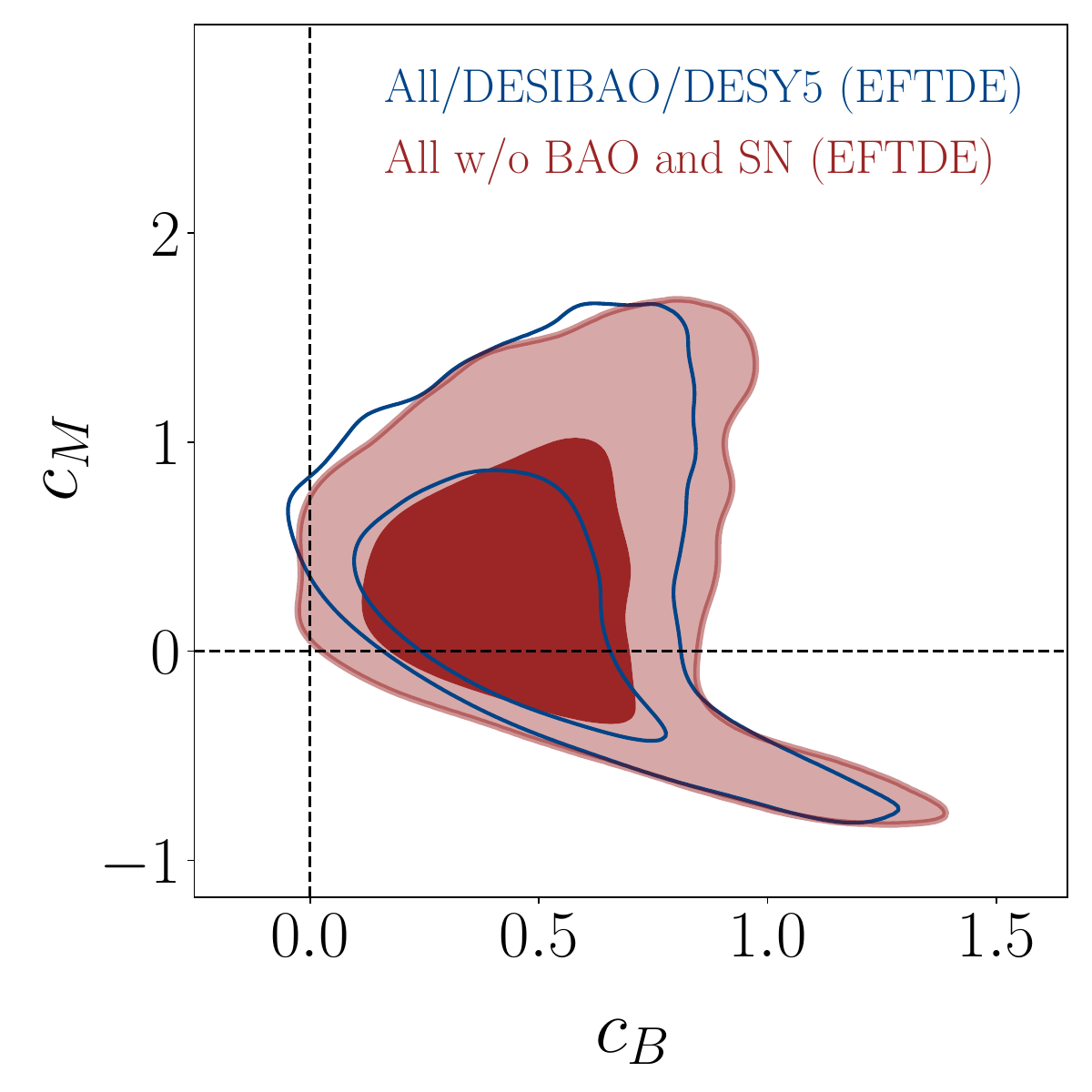}
    \end{minipage}
    \caption{ 2D posterior distributions of $\{w_0, \, w_a \}$ and $\{c_B, \, c_M \}$ for several variations of BAO and supernovae data.
    \textit{Top:} We compare the All dataset (with ext-BAO) with the All dataset in which DES Y5 has been replaced by Pantheon+.
    \textit{Bottom:}  We compare the All dataset and the All dataset in which DESI BAO DR2 and supernovae (SN) data have been removed.}
    \label{fig:noSupernovaenoBAO}
\end{figure}

In this section, we study the impact of the BAO and supernovae data in our analysis.
First, given that the preference for evolving dark energy depends on the choice of the supernovae and BAO likelihoods (see, \textit{e.g.},~\cite{DESI:2025zgx}), we carry out the ``All'' analysis by replacing DESI DR2 BAO with ext-BAO and DES Y5 with Pantheon+~\cite{Brout:2022vxf}, as shown in the top panel of Fig.~\ref{fig:noSupernovaenoBAO}. 
We observe that the preference for evolving dark energy decreases from $4.2\sigma$ (with DESI DR2 BAO and DES Y5) to $3.5 \sigma$ (with ext-BAO and DES Y5), then to $3.0\sigma$ (with ext-BAO and Pantheon+).
Regarding the EFTofDE parameters, we obtain very similar constraints when considering the ext-BAO and Pantheon+ likelihoods, with $c_B =0.46^{+0.18}_{-0.24} $ and $c_M =0.29^{+0.37}_{-0.51}$, due to the low correlations between the EFTofDE parameters and the background parameters (see Fig.~\ref{fig:baseline_all_w0wa_mg}). 

\begin{figure*}
    \centering
    \includegraphics[width=0.7\linewidth]{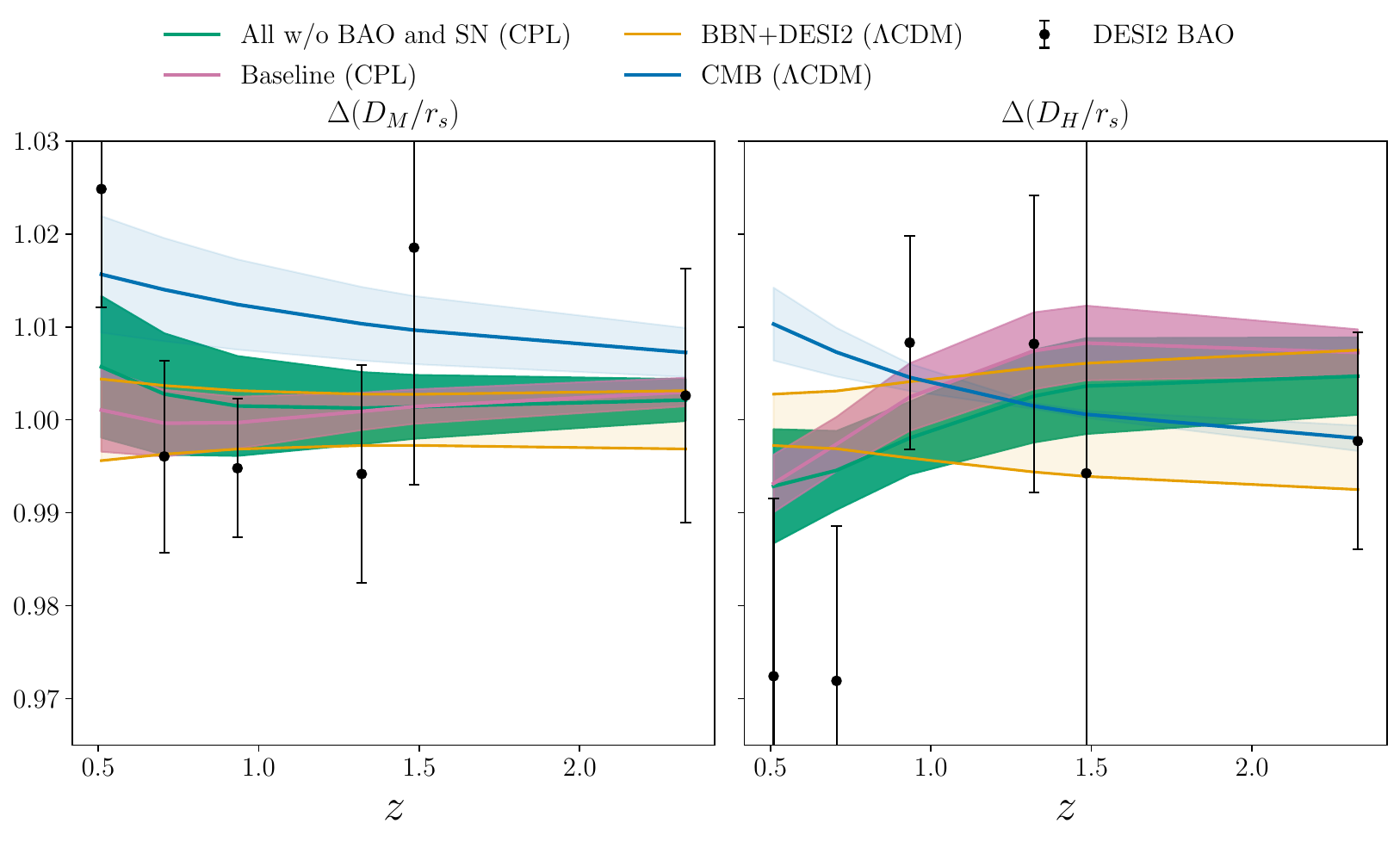}
    \caption{Residuals of the DESI DR2 BAO transverse $D_M/r_s$ (left) and radial $D_H/r_s$ (right) modes, together with theoretical predictions from different data and models, namely, DESI DR2 BAO (with BBN) under $\Lambda$CDM, \textit{Planck} PR4 under $\Lambda$CDM, Baseline under CPL (with PPF), and All without BAO and supernovae data under CPL (with PPF).
    The data and theoretical predictions are normalized to the prediction from DESI DR2 BAO + BBN (under $\Lambda$CDM).
    The shaded bands represent the $68 \%$ CL reconstructed from the full posterior distribution.} 
    \label{fig:BAO_data_theory}
\end{figure*}

Second, to gauge the impact of the BAO and supernovae likelihoods on our conclusions, we remove those datasets in our All analysis, as shown in Fig.~\ref{fig:noSupernovaenoBAO}.
For \textit{Planck} PR4 + Lensing + EFTBOSS + ISWL + DESI$C_\ell$, we obtain a $2.3 \sigma$ preference for the CPL model with PPF.
We also note that when we remove the BAO and supernovae data, the constraints on $c_B$ and $c_M$ remain unchanged (see Fig.~\ref{fig:noSupernovaenoBAO}).
In Fig.~\ref{fig:BAO_data_theory}, we present the DESI DR2 BAO data alongside the prediction from our analysis which does not include BAO and supernova data.
When we fit this prediction to the DESI DR2 BAO measurements, we obtain $\chi^2_{\rm min}({\rm BAO}) = 8.06$, which is significantly better than the  \textit{Planck} $\Lambda$CDM best-fit [with $\chi^2_{\rm min}({\rm BAO})=21.74$] or even to the $\Lambda$CDM fit to the DESI DR2 BAO + BBN dataset [with $\chi^2_{\rm min}({\rm BAO})=10.28$], as shown in Fig.~\ref{fig:BAO_data_theory}.
We conclude that our All analysis without BAO and supernovae data shows a slight preference for evolving dark energy and is able to better accommodate the DESI DR2 BAO measurements than $\Lambda$CDM.

\section{Conclusion}\label{sec:conclusion}
\label{sec:concl}

In this work, we constrain the CPL parametrization of the DE equation of state, considering either the parametrized post-Friedmann framework or the EFTofDE framework to describe the DE perturbations.
Within the EFTofDE framework, we vary $\alpha_B$, corresponding to the braiding
parameter, and $\alpha_M$, corresponding to the running of the Planck mass, both parametrized by $\alpha_i(a) = c_i \cdot \, \Omega_{\rm DE}(a)$.
The main goal of this paper is to constrain the parameter space of $\{w_0, \, w_a, \, c_B, \, c_M \}$ with (i) the EFTofLSS full-shape analysis of the power
spectrum and bispectrum of BOSS luminous red
galaxies (dubbed ``EFTBOSS''), (ii) the angular power spectrum $C_\ell^{T \kappa}$ from the cross-correlation between the \textit{Planck} PR4 temperature and lensing maps (dubbed ``ISWL''), and (iii) the auto-angular galaxy power spectra $C_\ell^{gg}$ from DESI luminous red galaxies, together with the cross-angular power spectra $C_\ell^{\kappa g}$ and $C_\ell^{T g}$ from the cross-correlation with the \textit{Planck} PR4 lensing and temperature maps (dubbed ``DESI$C_\ell$'').

One of the main novelties of this work is the measurement of $C_\ell^{T g}$ (in the DESI$C_\ell$ likelihood), detected with a signal-to-noise ratio of $2.04$, and the inclusion of the scales $20 < \ell < 79$ in $C_\ell^{gg}$, thanks to the \texttt{Swift$C_\ell$} code allowing us to go beyond the Limber approximation.

We summarize our main results in the following:
\begin{itemize}
    \item For the baseline analysis--including the primary power spectra and CMB gravitational lensing spectrum from \textit{Planck} PR4, the BAO measurements from DESI DR2, and the supernovae compilation from DES Y5--we obtain a preference for evolving dark energy over $\Lambda$ at $3.8 \sigma$ ($3.4 \sigma$) when considering the PPF (EFTofDE) framework. We also show that the choice of the CMB likelihood, for both the lensing and the primary spectra, can lead to shifts up to $0.4 \sigma$ in the reconstructed DE parameters.
    \item When we include the EFTBOSS likelihood in addition to the baseline dataset, we increase the preference for evolving dark energy to $4.7 \sigma$ ($4.3 \sigma$) for the PPF (EFTofDE) parametrization.
    The constraint on the 2D plane $\{w_0, \, w_a \}$ is improved by $\sim 30 \%$. Moreover, the EFTBOSS likelihood introduces a negative correlation between the EFTofDE parameters with a strong constraint on the upper bound of $c_M$, excluding $c_M \gtrsim 1.5$ at $98 \%$ CL.
    \item When we include the ISWL likelihood in addition to the baseline dataset, the preference for evolving dark energy and the constraint on the $\{w_0, \, w_a \}$ plane remain unchanged. However, the constraint on the $\{c_B, \, c_M \}$ plane is significantly improved by $\sim 50 \%$, while the ISWL likelihood induces a positive correlation between the EFT parameters with a strong constraint on the upper bound of $c_B$, excluding $c_B \gtrsim 1$ at $98 \%$ CL.
    \item When we include the DESI$C_\ell$ likelihood in addition to the baseline dataset, the preference for evolving dark energy and the constraint on the $\{w_0, \, w_a \}$ plane remain unchanged as well.
    However, the DESI$C_\ell$ likelihood leads to a negative correlation between the EFTofDE parameters (coming from $C_\ell^{gg}$), while improving the upper bound on $c_M$, excluding $c_M \gtrsim 2$ at $98 \%$ CL.
    \item When we combine all these likelihoods, we reduce the uncertainty on the $\{w_0, \, w_a \}$ plane by $\sim 50 \%$, and we increase the preference for evolving dark energy to $4.6 \sigma $ ($4.2 \sigma$) for the PPF (EFTofDE) framework.
    Regarding the EFTofDE parameters, we can exploit the various correlations presented above to significantly improve the constraint on the $\{ c_B, \, c_M \}$ parameter space by $177\%$, with $c_B = 0.46^{+0.16}_{-0.22}$ and $c_M = 0.31^{+0.39}_{-0.49}$.
    Those constraints are compatible with GR at $2\sigma$, and this conclusion does not change when we consider a $\Lambda$CDM background instead.
    \item Finally, when we remove the BAO and supernovae data, we obtain a hint for evolving dark energy at $2.3 \sigma$. We also note that the constraints on the EFTofDE parameters do not depend on the BAO and supernovae likelihoods (due to the low correlation between the background and EFToDE parameters).
\end{itemize}

There are several ways in which our work can be improved and extended. 
First, our analysis considers a linear galaxy bias expansion in the DESI$C_\ell$ likelihood, restricting our analysis of $C_\ell^{gg}$, $C_\ell^{\kappa g}$ and $C_\ell^{Tg}$ to linear scales. Incorporating a consistent EFTofLSS modeling for the galaxy overdensity field, as in Ref.~\cite{DAmico:2025zui}, or a hybrid-EFT approach, as in Ref.~\cite{Maus:2025rvz}, would allow the inclusion of smaller scales and further enhance the constraining power of the DESI$C_\ell$ likelihood. 
Second, developing a self-consistent covariance model between $C_\ell^{\kappa \kappa}$ and $C_\ell^{\kappa g}$ would allow us to include the low-$\ell$ CMB lensing data in the combined analysis and would help to further break parameter degeneracies.
Finally, our analysis can be reproduced for future high-precision galaxy maps, coming from DESI~\cite{DESI:2024jxi}, \textit{Euclid}~\cite{Euclid:2024yrr}, or LSST~\cite{LSST:2008ijt}, to improve the signal-to-noise ratio of the EFTBOSS and DESI$C_\ell$ likelihoods, and to enhance the measurement of the ISW signal, which is currently limited by observational noise.

\textit{Note added}. Recently, Ref.~\cite{Reeves:2025xau} performed a similar $5\times2$pt analysis. Their study focuses on early and late dark energy models, without considering the EFTofDE framework. Our results are consistent with, and independent of, theirs.

\begin{acknowledgments}
We thank Chi Zhang and Pierre Zhang for valuable suggestions and discussions.
ZL acknowledges the hospitality of the Galileo Galilei Institute for Theoretical Physics, the University of Montpellier, the University of Milan ``La Statale,'' as well as CERN during the completion of this work. ZL also thanks the INFN for partial support.
This work was supported in part by the National Key R\&D Program of China (2021YFC2203100), by the National Natural Science Foundation of China (12433002, 12261131497), by CAS young interdisciplinary innovation team (JCTD-2022-20), by 111 Project (B23042), by CSC Innovation Talent Funds, by USTC Fellowship for International Cooperation, and by USTC Research Funds of the Double First-Class Initiative.
TS acknowledges the European Union’s Horizon Europe research and innovation program under the Marie Skłodowska-Curie Staff Exchange grant agreement No 101086085 – ASYMMETRY.
This work was partially supported by the computational resources from the  LUPM’s cloud computing infrastructure founded by Ocevu labex and France-Grilles.
We finally acknowledge the use the standard \texttt{python} packages, such as \texttt{numpy}~\cite{harris2020array}, \texttt{scipy}~\cite{2020SciPy-NMeth}, and \texttt{matplotlib}~\cite{Hunter:2007}.
\end{acknowledgments}

\appendix

\section{Details and validation of the DESI$C_\ell$ likelihood}\label{appendix:mapar}

\subsection{Details on the DESI$C_\ell$ likelihood}

\begin{figure}[]
    \centering
    \begin{minipage}{0.5\textwidth}
        \centering
         \includegraphics[width=1.\textwidth]{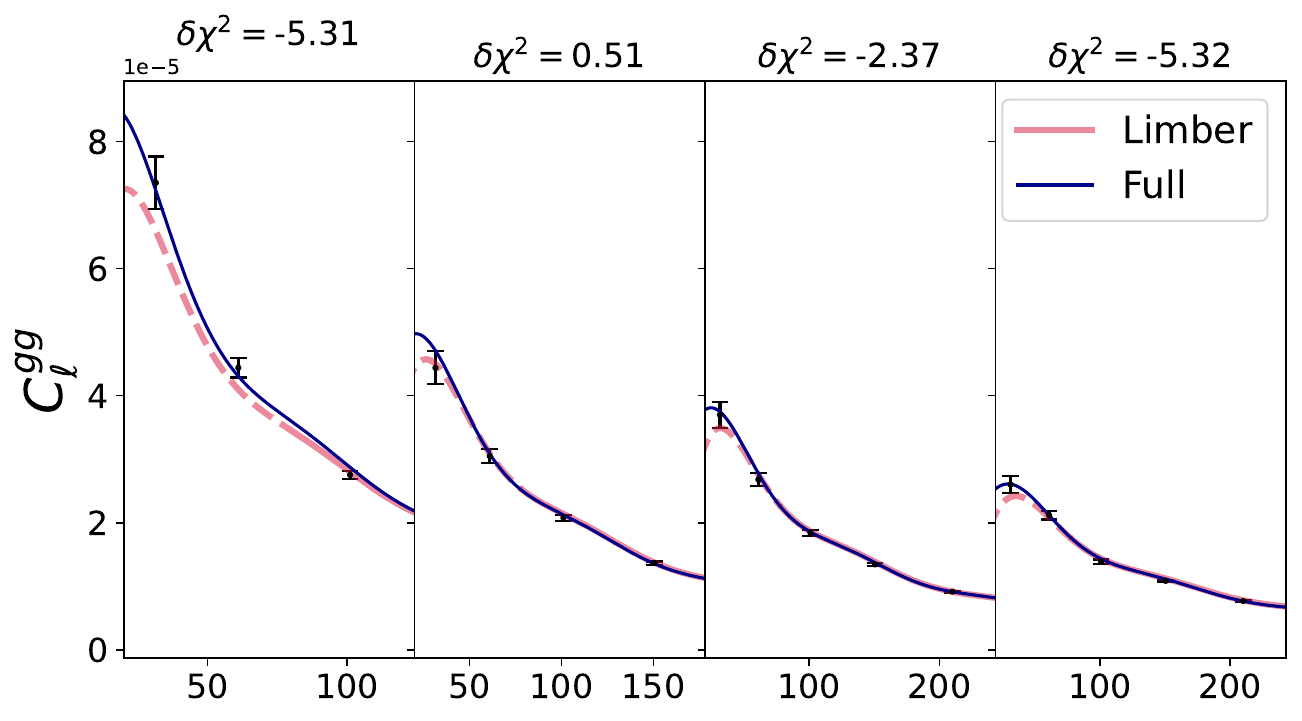}
         \includegraphics[width=1.\textwidth]{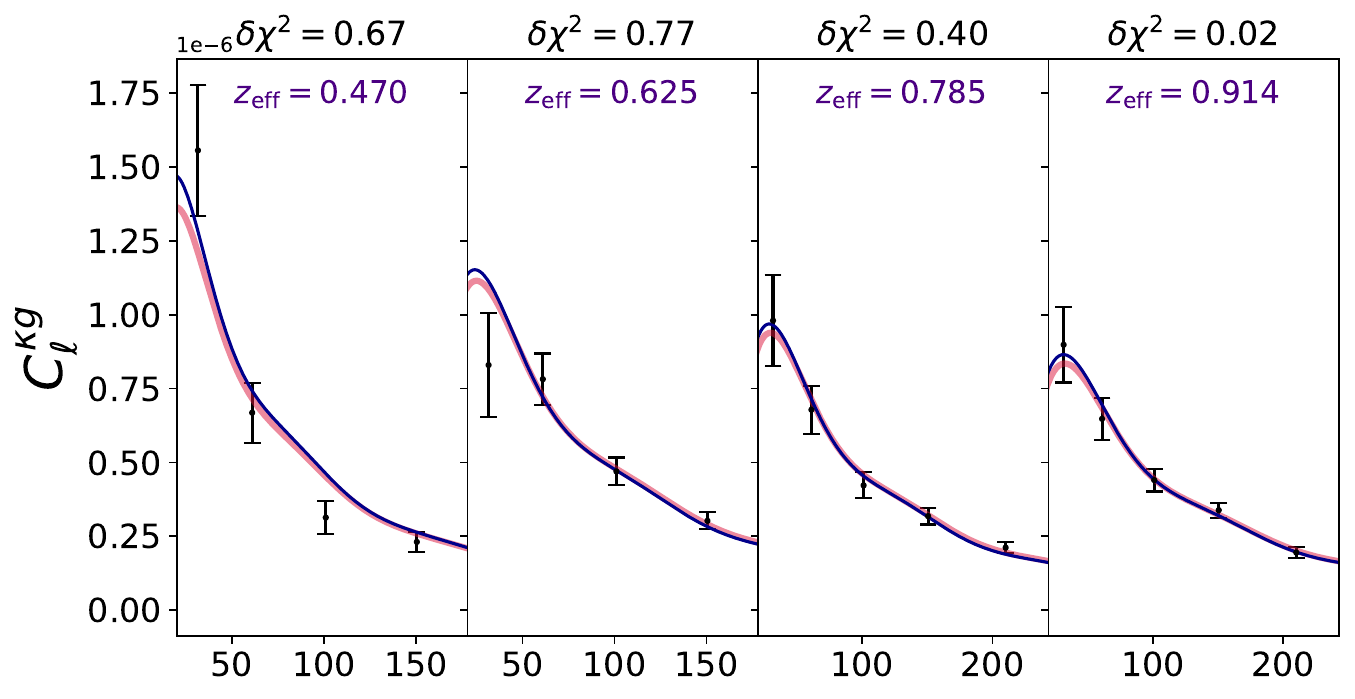}
         \includegraphics[width=1.\textwidth]{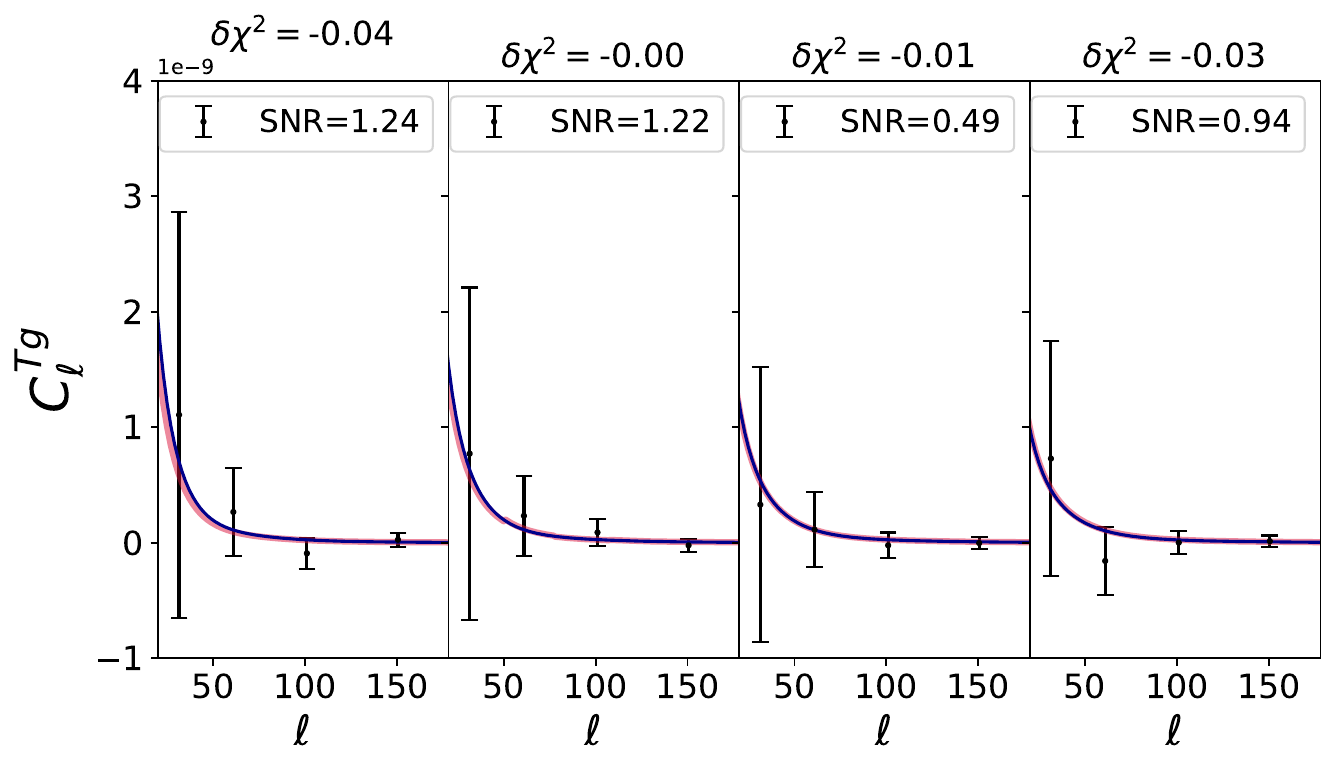}
    \end{minipage}
    \caption{The $C_\ell^{gg}$, $C_\ell^{\kappa g}$, and $C_\ell^{Tg}$ data points considered in the DESI$C_\ell$ likelihood, together with the best-fit prediction from the full integral (dark blue) or from the Limber approximation (crimson). The dashed lines correspond to the regimes where the Limber approximation fails. We also display the individual $\delta \chi^2 = \chi^2({\rm Limber}) - \chi^2({\rm full \ integral})$ for each angular power spectra and each redshift bin.}.
    \label{fig:namaster_cltg}
\end{figure}

This appendix documents the implementation of the DESI$C_\ell$ likelihood used in \texttt{MontePython} and highlights the differences with \texttt{Mapar}~\cite{Sailer:2024jrx}, a publicly available likelihood designed to extract cosmological information from the auto- and cross-correlations between CMB lensing and luminous red galaxies (LRGs) maps. This likelihood makes use of the \texttt{NaMaster} code~\cite{Alonso:2018jzx} to measure the angular power spectra $C_\ell^{\kappa g}$ and $C_\ell^{gg}$ as well as their covariance matrices under the Gaussian approximation. 
For the theoretical prediction, Ref.~\cite{Sailer:2024jrx} numerically integrates the angular power spectra using the Limber approximation. In addition, the \texttt{Mapar} likelihood considers a hybrid effective field theory (HEFT) model~\cite{Kokron:2021xgh} to determine the galaxy overdensity field from the underlying dark matter field.
In our \texttt{MontePython} likelihood (DESI$C_\ell$), we use the data and covariances from \texttt{Mapar} for both  $C_\ell^{\kappa g}$ and $C_\ell^{gg}$. However, we make use of \texttt{Swift$C_\ell$} for the theoretical prediction, allowing us to go beyond the Limber approximation. This enables us to include additional data in the range $20 \leq \ell \leq 79$ for $C_\ell^{gg}$, thus improving the constraints.
In addition, in our work, we focus on linear scales and adopt a linear galaxy bias model, meaning that we do not consider the small-scale data included in Ref.~\cite{Sailer:2024jrx}. 
In particular, this reference includes data up to $l_{\rm max} = 600$ for $C_\ell^{gg}$ and $C_\ell^{\kappa g}$, while we perform our analysis up to $l_{\rm max} = 124 - 243$ (see Sec.~\ref{sec:angularspectramodelling}), as determined in Ref.~\cite{Sailer:2024jrx} for the linear bias galaxy model.
These scales correspond to the maximum scales such that the linear prediction does not deviate significantly from the nonlinear prediction given the data covariance~\cite{Planck:2015bue,DES:2022ccp,Zanoletti:2025xdc}.\footnote{ {Note that Ref.~\cite{Sailer:2024jrx} did not include the first redshift bin for $C_\ell^{gg}$ and $C_\ell^{\kappa g}$ in their linear bias analysis due to the limited scale range probed by the linear perturbation theory at low redshift. Given that in our analysis we go to higher scales, the constraining power of the first redshift bin becomes relevant, and we determine $\ell_{\rm max}$ using the procedure describe in the main text.}} 
To do so, the $\chi^2$ obtained using the linear perturbation theory prediction is compared to the one obtained using the \texttt{Halofit} prediction~\cite{Smith:2002dz,Takahashi:2012em} for several $\ell_{\rm max}$, and we impose that $\chi^2 = \Delta C_\ell^\mathrm{T} \, \mathbf{C}^{-1} \, \Delta C_\ell<1.0$, where
$\Delta C_\ell = C_\ell^{\mathrm{lin}} - C_\ell^{\mathrm{nl}}$.
We further adopt the more conservative choice compared to Table 2 of Ref.~\cite{Sailer:2024jrx}.

We adopt the same pixel and bandpower window functions as those used in the \texttt{Mapar} likelihood. 
The theoretical angular spectra are multiplied by the pixel window function to correct for the pixelization introduced when constructing a \texttt{Healpix}~\cite{Zonca2019,2005ApJ...622..759G} overdensity map from the galaxy catalog, 
 and then convolved with the bandpower window function to match the mask-deconvolved measurements from \texttt{NaMaster}.
Finally, we note that the shot noise parameters are not analytically marginalized, but are instead varied as free parameters in the MCMC sampling.

One of the main novelties of our work is that we additionally include the cross-correlation between LRGs and \textit{Planck} PR4 temperature maps (SEVEM) as shown in Fig.~\ref{fig:namaster_cltg}. To measure $C_\ell^{Tg}$, we use \texttt{NaMaster}, which provides a pseudo-$C_\ell$ estimator that accounts for survey masks. To cross-check our results, we also measure $C_\ell^{Tg}$ with the \texttt{Polspice} code~\cite{2011ascl.soft09005C}, and obtain a good consistency up to $0.2\sigma$ for all data points. 
To determine the maximum scale $\ell_{\rm max}$, we use the procedure detailed before, allowing us to determine $\ell_{\rm max} = 2212$.
However, due to the limited statistical power of the small-scale modes, we fix $\ell_{\rm max}=178$ for $C_\ell^{Tg}$ across the four redshift bins.
Finally, the measurements of $C_\ell^{Tg}$ yield a cumulative signal-to-noise ratio~\cite{Sailer:2024jrx,Kim:2024dmg} of ${\rm SNR}=\sqrt{\sum_i\chi^2(C_\ell^{Tg,i})}=2.04$.

In Fig.~\ref{fig:namaster_cltg}, we display the data points considered in this work for $C_\ell^{gg}$, $C_\ell^{\kappa g}$, and $C_\ell^{Tg}$, together with the best-fit predictions either from the Limber approximation or from the full integral (computed with \texttt{Swift$C_\ell$}).\footnote{We note that the Limber prediction for the low-$\ell$ part of the galaxy auto-angular power spectrum is extrapolated from the small-scale prediction (dashed lines in Fig.~\ref{fig:namaster_cltg}).}
We also show the individual $\delta \chi^2 =\chi^2({\rm full \ integral})- \chi^2({\rm Limber})$ for each angular power spectra and each redshift bin.
The largest improvements are found in the first and fourth redshift bins of $C_\ell^{gg}$, where the $\chi^2$ differences reach $-5.31$ and $-5.32$, respectively. 
For $C_\ell^{\kappa g}$ and $C_\ell^{Tg}$, the difference between the full integral and the Limber approximation is negligible, even at large scales.
This is consistent with Ref.~\cite{Sailer:2024jrx}, which applies the Limber approximation until the same $\ell_{\rm min}$ as in our analysis for $C_\ell^{\kappa g}$ but not for $C_\ell^{gg}$.\\

\subsection{Validation of our pipeline}

\begin{figure}[!h]
    \centering
    \begin{minipage}{0.5\textwidth}
        \centering
         \includegraphics[width=1.\textwidth]{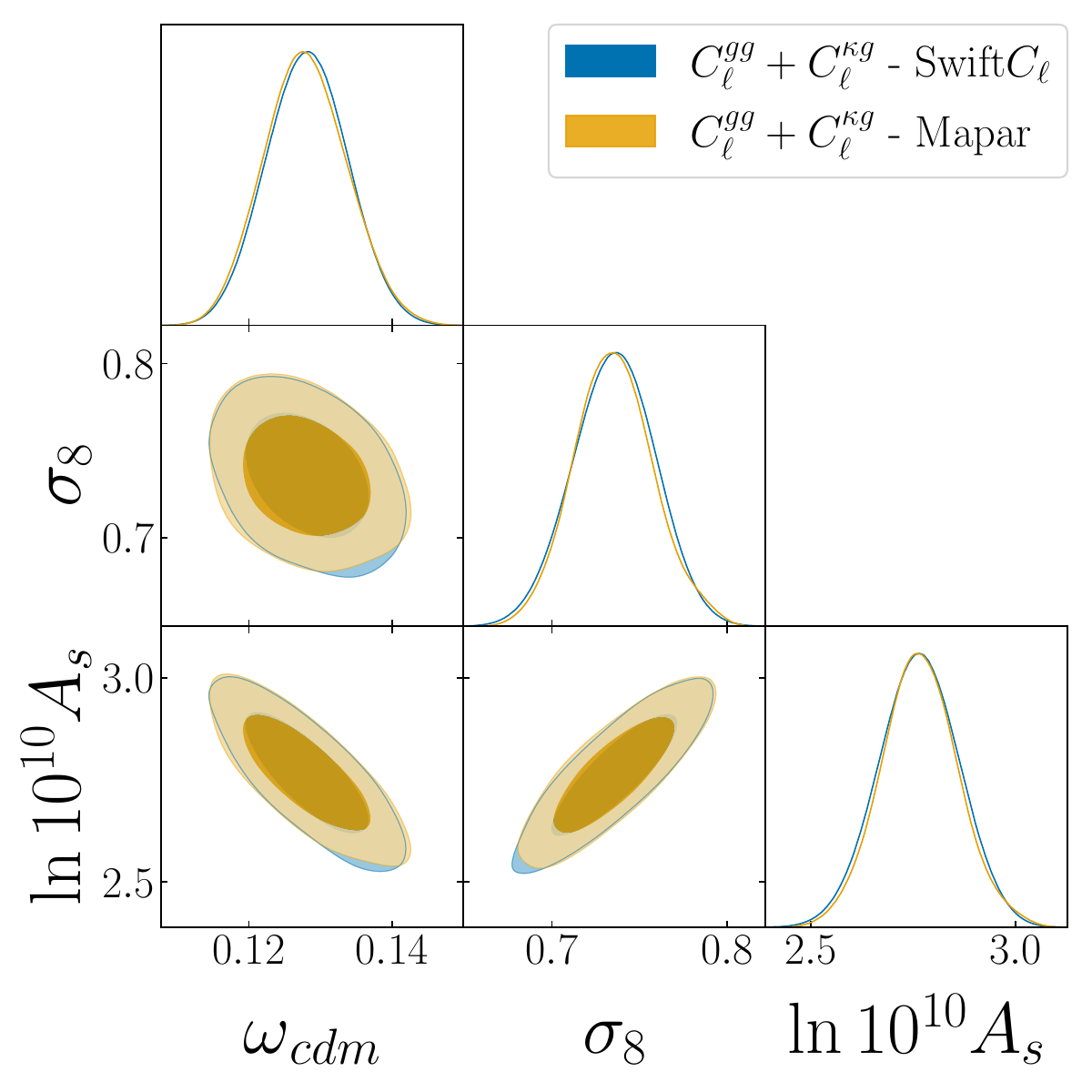}
    \end{minipage}
    \caption{1D and 2D posterior distributions reconstructed from our DESI$C_\ell$ likelihood (using \texttt{MontePython} and \texttt{Swift$C_\ell$}) and from the \texttt{Mapar} likelihood (using \texttt{Cobaya} and the Limber approximation) for the $\Lambda$CDM model.}
    \label{fig:compare_mapar_swift}
\end{figure}

To validate our pipeline, we compare in  Fig.~\ref{fig:compare_mapar_swift} the cosmological posteriors from our DESI$C_\ell$ analysis, using \texttt{MontePython} and \texttt{Swift$C_\ell$}, with those from \texttt{Mapar}, using \texttt{Cobaya} and the Limber approximation.
For comparison purposes, we include only $C_\ell^{gg}$ and $C_\ell^{\kappa g}$, while we consider the scale cuts, determined in Ref.~\cite{Sailer:2024jrx}, which are valid for the Limber approximation and the linear bias model across the four redshift bins. We further note that the shot noise terms are fixed to their fiducial values. 
The excellent agreement between the two approaches confirms both the validity of our \texttt{MontePython} likelihood implementation and the validity of the \texttt{Swift$C_\ell$} calculation.

\subsection{Assessment of the constraining power of the DESI$C_\ell$ likelihood}\label{sec:compare_fullintegral_limber}

\begin{figure}[!h]
    \centering
    \begin{minipage}{0.5\textwidth}
        \centering
         \includegraphics[width=1.\textwidth]{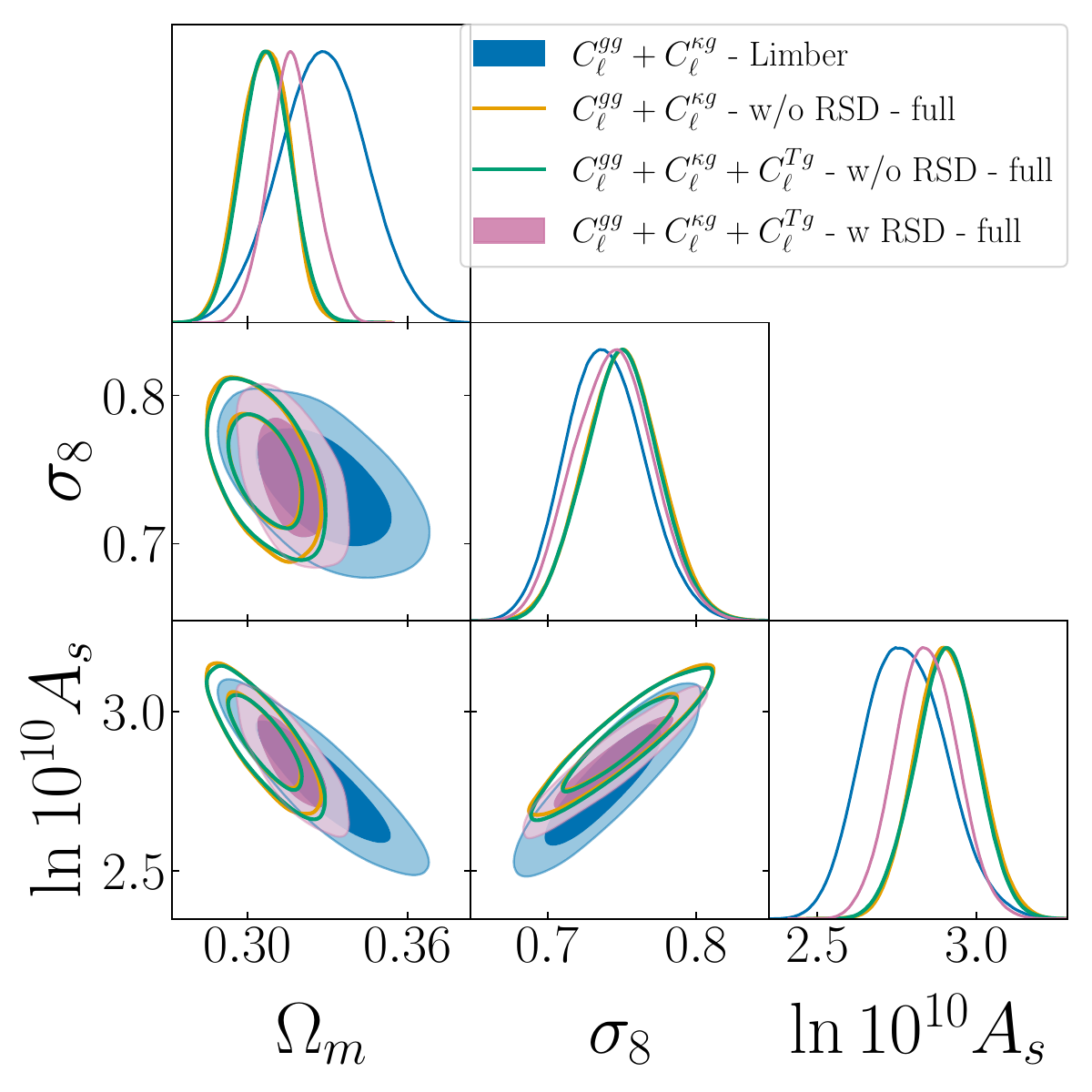}
    \end{minipage}
    \caption{1D and 2D posterior distributions reconstructed from our DESI$C_\ell$ likelihood for the $\Lambda$CDM model and for several analysis settings. In particular, we evaluate the impact of (i) the full integral prediction compared to the Limber approximation, (ii) the RSD correction, and (iii) the inclusion of the $C_\ell^{Tg}$ data.
    }
    \label{fig:compare_limber_full}
\end{figure}

\begin{figure*}[htbp]
    \centering
    \begin{minipage}{1\textwidth}
        \centering
        % ======= Legend above =======
    \small
    \fbox{%
    {%
    \begin{tabular}{@{}c@{}}
    \textcolor[HTML]{0072B2}{$C_\ell^{gg}+C_\ell^{\kappa g}$ -- Limber} \quad
    \textcolor[HTML]{E69F00}{$C_\ell^{gg}+C_\ell^{\kappa g}$ -- w/o RSD -- full} \\[0.4em]
    \textcolor[HTML]{009E73}{$C_\ell^{gg}+C_\ell^{\kappa g}+C_\ell^{Tg}$ -- w/o RSD -- full} \quad
    \textcolor[HTML]{CC79A7}{$C_\ell^{gg}+C_\ell^{\kappa g}+C_\ell^{Tg}$ -- w RSD -- full}
    \end{tabular}%
    }%
    }
    \vspace{0.5em} % 控制legend与图像的垂直间距
         \includegraphics[width=0.45\textwidth]{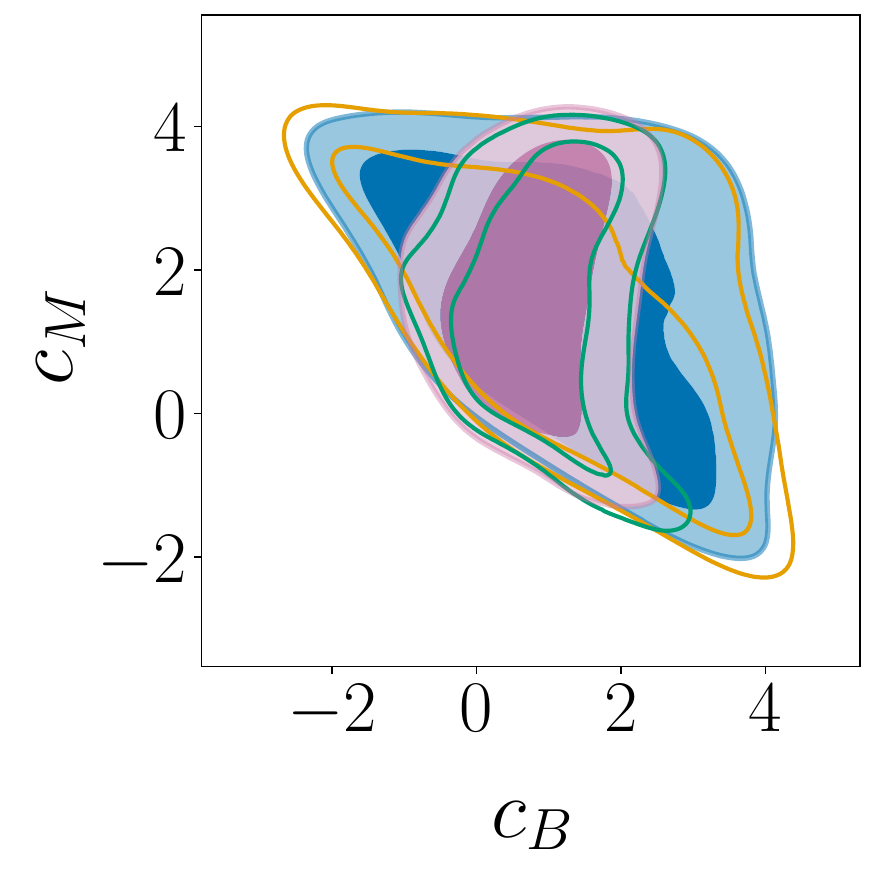}
         \includegraphics[width=0.45\textwidth]{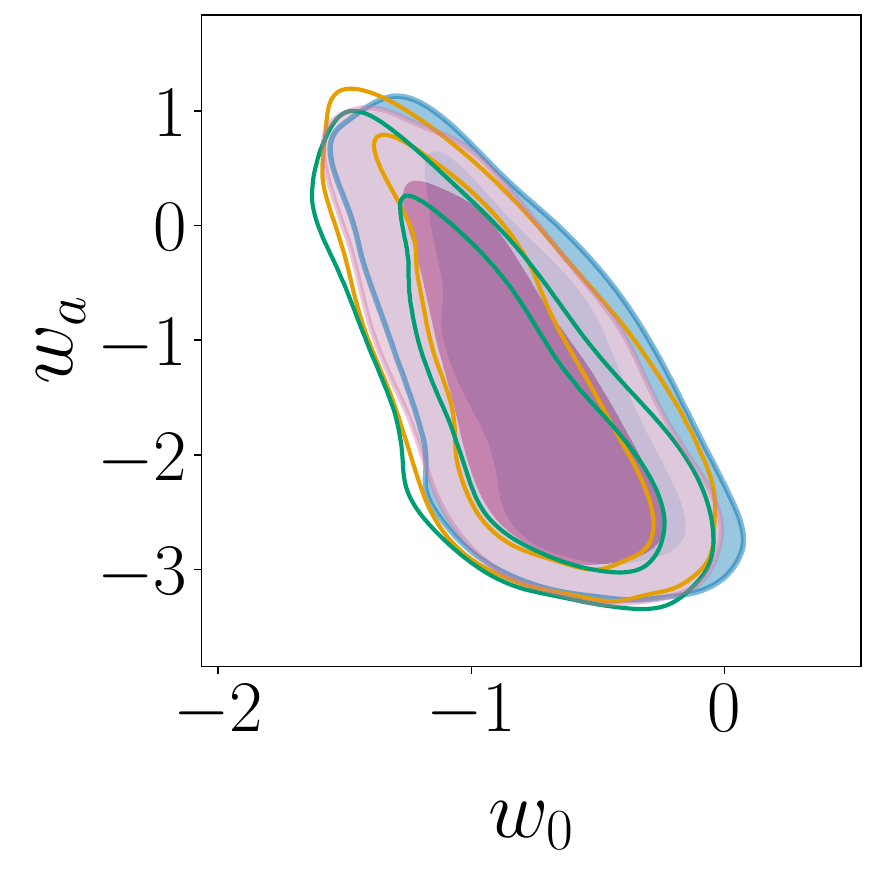}
    \end{minipage}
    \caption{
    2D posterior distributions of $\{c_B, \, c_M \}$ and $\{w_0, \, w_a \}$ reconstructed from our DESI$C_\ell$ likelihood for our EFTofDE configuration and for several analysis settings. In particular, we evaluate the impact of (i) the \texttt{Swift$C_\ell$} prediction compared to the Limber approximation, (ii) the RSD correction and (iii) the inclusion of the $C_\ell^{Tg}$ data.
    }
    \label{fig:ggkgtg_alpha_w0wa}
\end{figure*}

In Fig.~\ref{fig:compare_limber_full}, we compare the cosmological constraints obtained for the $\Lambda$CDM model with different analysis settings within our DESI$C_\ell$ likelihood.
We vary here two cosmological parameters, namely, $\{\omega_{\rm cdm}, \, \ln(10^{10} A_s)\}$, and twelve nuisance parameters, namely, $\{b_i, \, s_{\mu,i}, \,  \mathrm{SN}_i\}$, where $i=1,\dots,4$ corresponds to the four redshift bins. In Fig.~\ref{fig:ggkgtg_alpha_w0wa}, we show the equivalent analysis for our EFTofDE configuration.

First, we evaluate the impact of the inclusion of the additional data points in the range $20 < \ell < 79$ for $C_\ell^{gg}$ (see Limber vs full). 
With the full integral analysis, we obtain $\Omega_m = 0.3063 \pm 0.0090$, $\sigma_8 = 0.750 \pm 0.025$, ${\rm ln}10^{10}A_{s } = 2.908\pm 0.098$ and $S_8=0.7576\pm0.022$. This respectively corresponds to an improvement in the constraining power of $43.8 \%$, $3.8 \%$, $21.6 \%$, $5.5 \%$, and a shift of $1.7 \sigma$, $0.4 \sigma$, $1.1 \sigma$, $0.6 \sigma$ over the Limber approximation analysis. 
However, the inclusion of these additional data points does not lead to a noticeable improvement in the constraints on the DE parameters (see Fig.~\ref{fig:ggkgtg_alpha_w0wa}).

Second, we evaluate the impact of the inclusion of the $C_\ell^{Tg}$ data in the DESI$C_\ell$ likelihood, an important novelty compared to the \texttt{Mapar} likelihood of Ref.~\cite{Sailer:2024jrx}.
Although the inclusion of $C_\ell^{Tg}$ does not improve the constraints on the standard $\Lambda$CDM parameters (see Fig.~\ref{fig:compare_limber_full}), it does provide additional constraining power on the EFToDE parameters $c_B$ and $c_M$, as illustrated in Fig.~\ref{fig:ggkgtg_alpha_w0wa}. 
The constraints on $c_B$ and $c_M$ are improved by $55\%$ and $21\%$, from $c_B=1.2^{+2.0}_{-1.5}$ and $c_M = 1.4^{+2.3}_{-1.5}$ (without $C_\ell^{Tg}$) to $c_B=0.94\pm 0.78$ and $c_M = 1.5^{+1.7}_{-1.3}$ (with $C_\ell^{Tg}$). 
This is attributed to the improved sensitivity to the Weyl potential thanks to the temperature-galaxy cross-correlation. 
Concerning the constraints on the dark energy equation of state parameters $w_0$ and $w_a$, we do not get much improvement.

% colors = ['#0072B2','#E69F00', '#009E73', '#CC79A7']

Third, always in Fig.~\ref{fig:compare_limber_full}, we gauge the impact of adding the redshift-space distortion (RSD) contribution, which is significant only on the largest angular scales ($\ell \lesssim 79$~\cite{Sailer:2024jrx}). We observe a noticeable shift in the parameter posterior of $1.1 \sigma$, $0.2 \sigma$, and $0.6 \sigma$ for $\Omega_m$, $\sigma_8$, and $\ln(10^{10}A_s)$.
We then obtain as a final constraint $\Omega{}_{m } = 0.3167\pm 0.0088$, $\sigma_8 = 0.744\pm 0.026$, $ln10^{10}A_{s } = 2.842\pm 0.097$, and $S_8=0.76\pm 0.024$ for $C_\ell^{gg}+C_{\ell}^{\kappa g}+C_\ell^{Tg}$ (with the full integral and the RSD contribution).
For the DE parameters, we do not observe any significant shift and improvement, and we obtain as a final constraint $w_0 = -0.79\pm 0.32$,  $w_a = -1.46^{+0.61}_{-1.4}$, $c_B= 0.82\pm 0.78$, and $c_M = 1.7\pm 1.3$.

\section{Supplementary analysis products} \label{app:supp_results}

\begin{figure*}
    \centering
    \includegraphics[width=1\linewidth]{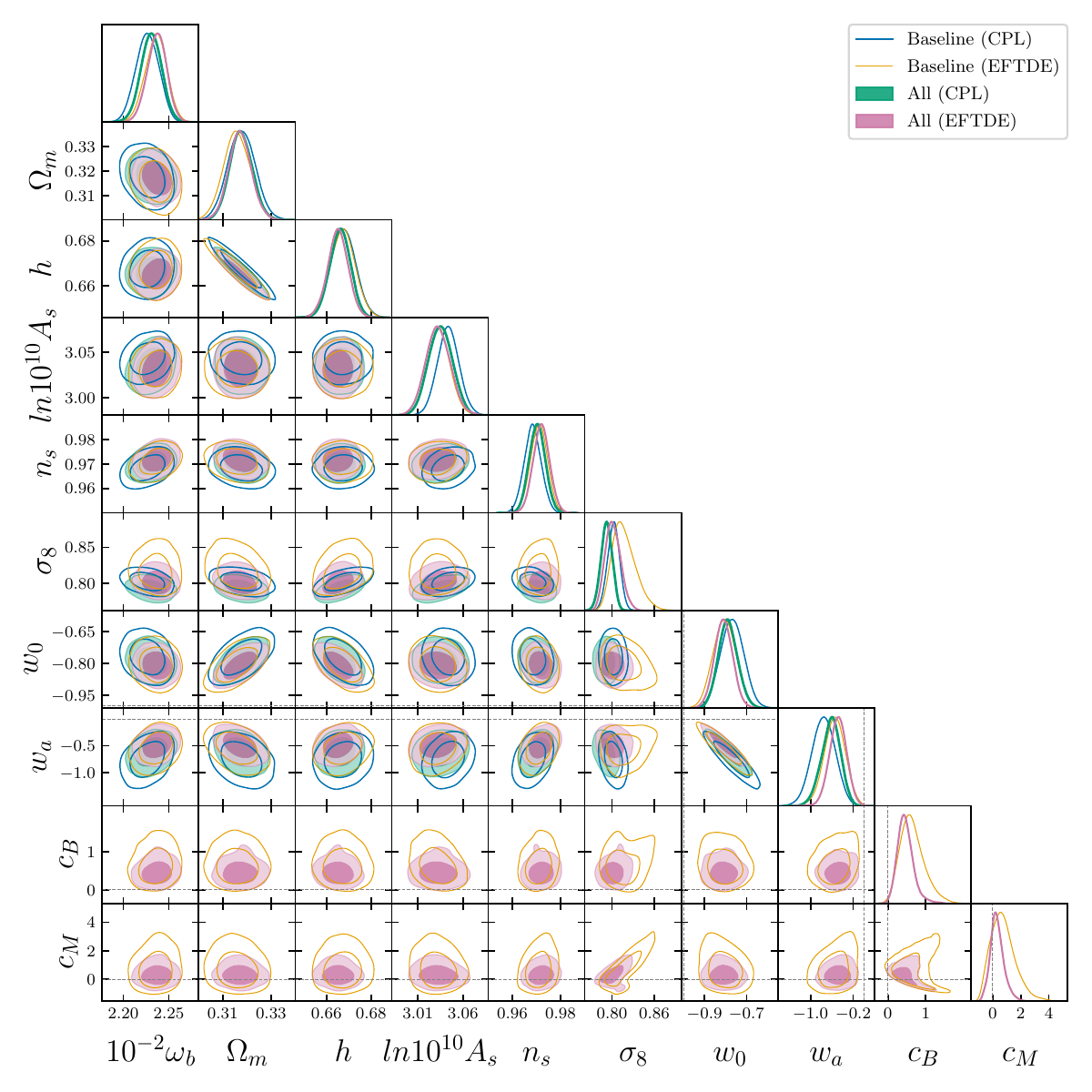}
    \caption{1D and 2D posterior distributions reconstructed from our Baseline and All analyses for the CPL model with and without EFTofDE. }
    \label{fig:baseline_all_w0wa_mg}
\end{figure*}

In this appendix, we report the cosmological constraints, the $\Delta \chi_{\rm min}^2$ with respect to $\Lambda$CDM, and the preference over $\Lambda$CDM, for the CPL model with PPF in Tab.~\ref{tab:bestfit_params_cpl} and for the CPL model with EFTofDE in  Tab.~\ref{tab:bestfit_params_eftde}.
In addition, in Tab.~\ref{tab:bestfit_params_eftde_transposed}, we display the best-fit $\chi^2$ per experiment reconstructed from our All analysis for the $\Lambda$CDM model, the CPL model with PPF, and the CPL model with EFTofDE.
Finally, in Fig.~\ref{fig:baseline_all_w0wa_mg}, we display the 1D and 2D posterior distributions reconstructed from our Baseline and All analyses for the CPL model with and without EFTofDE.

\begin{table*}[!ht]
    \centering
    \begin{tabular}{|c|c|c|c|c|c|c|}
        \hline
        CPL & Baseline &Baseline + ISWL & Baseline + EFTBOSS & Baseline + DESI$C_\ell$ & All& All w/o BAO and SN\\
        \hline\hline
        $10^2\omega_b$ 
        &  2.233 &2.229&2.226&2.228&2.230&2.230\\
        &$2.226\pm 0.012$ &$2.226\pm 0.012$&$2.229^{+0.013}_{-0.012}$&$2.230\pm 0.012$&$2.231\pm 0.011$&$2.230^{+0.013}_{-0.011}$\\
        \hline
        $\Omega_m$ 
        & 0.3173 &0.3181&0.3177&0.3170&0.3174&0.3168\\
        & $0.3178\pm 0.0056$&$0.3176\pm 0.0055$&$0.3183\pm 0.0049$&$0.3172\pm 0.0055$&$0.3176\pm 0.0047$&$0.3218^{+0.0080}_{-0.0091}$\\
        \hline
        $h$ 
        & $0.6673$ &0.6664&0.6675&0.6671&0.6666&0.6680\\
        &$0.6675\pm 0.0056$&$0.6676\pm 0.0055$&$0.6661\pm 0.0047$&$0.6670\pm 0.0055$&$0.6661\pm 0.0046$&$0.6621\pm 0.0084$ \\
        \hline
        $\ln10^{10}A_s$ 
        & $3.046$ &3.044&3.044&3.041&3.042&3.039 \\
        &$3.044\pm 0.012$&$3.043\pm 0.012$&$3.042\pm 0.012$&$3.039^{+0.013}_{-0.014}$&$3.035\pm 0.013$&$ 3.035\pm 0.013$ \\
        \hline
        $n_s$ 
        & $0.9692$ &0.9693&0.9700&0.9712&0.9707&0.9697\\
        &$0.9684\pm 0.0035$&$0.9684\pm 0.0035$&$0.9693\pm 0.0036$&$0.9698^{+0.0038}_{-0.0033}$&$0.9703\pm 0.0033$&$0.9698\pm 0.0037$ \\
        \hline
        $\sigma_8$ 
        & $0.8003$&0.7993&0.8033&0.7981&0.7971&0.7993 \\
        &$0.8023\pm 0.0085$&$0.8022\pm 0.0083$&$0.7975\pm 0.0077$&$0.7964\pm 0.0090$&$0.7925\pm 0.0080$&$ 0.789\pm 0.010$ \\
        \hline
        $w_0$ 
        & $-0.788$ &-0.779&-0.778&-0.784&-0.778&-0.774\\
        &$-0.769\pm 0.056$&$ -0.768\pm 0.056$&$-0.780\pm 0.049$&$-0.780\pm 0.057$&$ -0.788\pm 0.046$&$ -0.777^{+0.078}_{-0.10}$ \\
        \hline
        $w_a$ 
        & $-0.65$&-0.68&-0.72&-0.66&-0.67&-0.72 \\
        &$-0.75\pm 0.22$&$-0.76^{+0.23}_{-0.21}$&$-0.67^{+0.21}_{-0.18}$&$-0.67^{+0.23}_{-0.20}$&$-0.62^{+0.19}_{-0.16}$&$-0.62^{+0.37}_{-0.27}$ \\

        \hline\hline
        $\chi^2_{\rm min}$
        &32218.2&32220.2&32863.6&32247.6&32895.2&31249.6\\
        \hline
        $\Delta\chi^2_{\rm min}$
        &-17.8&-18.0&-25.4&-17.0&-24.6&-7.6\\
        \hline
        n-$\sigma$
        &3.8&3.8&4.7&3.7&4.6&2.3
        \\
        \hline
    \end{tabular}
    \caption{Best-fit values (first row) and mean $\pm 1 \sigma$ (second row) of the reconstructed parameters for the CPL model with PPF. We also display the total best-fit $\chi^2_{\rm min}$, the $\Delta \chi_{\rm min}^2$ with respect to $\Lambda$CDM, and the preference over $\Lambda$CDM.}
    \label{tab:bestfit_params_cpl}
\end{table*}

\begin{table*}[!ht]
    \centering
    \begin{tabular}{|c|c|c|c|c|c|c|}
        \hline
        EFTofDE & Baseline &Baseline + ISWL & Baseline + EFTBOSS & Baseline + DESI$C_\ell$ & All& All w/o BAO and SN\\
        \hline\hline
        $10^2\omega_b$ 
        &  2.234&2.235&2.233&2.236&2.236&2.236  \\
        &$2.236\pm 0.012$&$2.235\pm 0.012$&$2.236^{+0.013}_{-0.011}$&$2.237\pm 0.011$&$2.237^{+0.011}_{-0.010}$&$2.235\pm 0.012$\\
        \hline
        $\Omega_m$ 
        &0.3165& 0.3166&0.3171&0.3167&0.3162&0.3173 \\
        & $0.3159\pm 0.0056$&$0.3158\pm 0.0056$&$0.3173\pm 0.0047$&$0.3159\pm 0.0054$&$0.3172\pm 0.0046$&$0.3234\pm 0.0091$\\
        \hline
        $h$ 
        &0.6675&0.6671&0.6672& 0.6666&0.6673&0.6661\\
        &$0.6673\pm 0.0056$& $0.6674\pm 0.0056$&$0.6655\pm 0.0047$&$0.6670\pm 0.0055$&$0.6652\pm 0.0045$&$ 0.6591\pm 0.0089$\\
        \hline
        $\ln10^{10}A_s$ 
        &3.037&3.038&3.037&3.037&3.040&3.039 \\
        &$3.032\pm 0.013$&$3.036\pm 0.013$&$3.029\pm 0.013$&$3.032\pm 0.014$&$3.033\pm 0.014$&$3.033\pm 0.014$ \\
        \hline
        $n_s$ 
        &0.9706&0.9711&0.9703&0.9715&0.9720&0.9725 \\
        &$0.9710\pm 0.0035$&$0.9710\pm 0.0035$&$0.9710\pm 0.0035$&$0.9715\pm 0.0033$&$0.9720\pm 0.0034$&$0.9716\pm 0.0036$ \\
        \hline
        $\sigma_8$ 
        &0.805& 0.807&0.804&0.801&0.805&0.803 \\
        & $0.817^{+0.013}_{-0.019}$&$ 0.821^{+0.014}_{-0.016}$&$0.8033^{+0.0087}_{-0.011}$&$0.812^{+0.011}_{-0.017}$&$ 0.8016^{+0.0097}_{-0.011}$&$0.796\pm 0.013$\\
        \hline
        $w_0$ 
        &-0.788& -0.792&-0.789&-0.792&-0.798&-0.805\\
        &$-0.805\pm 0.054$&$-0.809\pm 0.052$&$-0.801\pm 0.045$&$-0.798\pm 0.054$&$-0.809\pm 0.043$&$-0.820^{+0.078}_{-0.088}$ \\
        \hline
        $w_a$ 
        & -0.64&-0.61&-0.64&-0.59&-0.58 &-0.54\\
        & $-0.55\pm 0.20$&$-0.53\pm 0.18$&$-0.53\pm 0.17$&$-0.56^{+0.21}_{-0.18}$&$-0.49\pm 0.16$&$ -0.38\pm 0.28$\\
        \hline
        $c_B$ 
        & 0.36&0.20&0.60&0.52&0.27 &0.28\\
        & $0.65^{+0.24}_{-0.37}$&$0.45^{+0.16}_{-0.30}$&$ 0.67^{+0.23}_{-0.37}$&$0.81^{+0.28}_{-0.41}$&$0.46^{+0.16}_{-0.22}$&$ 0.50^{+0.18}_{-0.26}$\\
        \hline
        $c_M$ 
        & 0.10&0.26&-0.33&-0.03&0.18&0.19\\
        & $ 0.75^{+0.63}_{-1.0}$&$ 0.98^{+0.56}_{-0.82}$&$ 0.16^{+0.41}_{-0.60}$&$ 0.38^{+0.53}_{-0.83}$&$ 0.31^{+0.39}_{-0.49}$&$ 0.35^{+0.39}_{-0.54}$\\

        \hline\hline
        $\chi^2_{\rm min}$
        &32216.4&32218.8&32862.0&32245.2&32893.6&31247.8\\
        \hline
        $\Delta\chi^2_{\rm min}$
        &-19.6&-19.4&-27.0&-19.4&-26.2&-9.4\\
        \hline
        n-$\sigma$
        &3.4&3.4&4.3&3.4&4.2&1.9
        \\
        \hline
    \end{tabular}
    \caption{Best-fit values (first row) and mean $\pm 1 \sigma$ (second row) of the reconstructed parameters of the CPL model model with EFTofDE. We also display the total best-fit $\chi^2_{\rm min}$, the $\Delta \chi_{\rm min}^2$ with respect to $\Lambda$CDM, and the preference over $\Lambda$CDM.}
    \label{tab:bestfit_params_eftde}
\end{table*}

\begin{table*}[!ht]
    \centering
    \begin{tabular}{|c|c|c|c|c|c|c|c|c|c|c|}
        \hline
        $\chi^2_{\rm min}$ & high-$\ell$ TTTEEE & low-$\ell$ TT & low-$\ell$ EE & Lensing & ISWL & DESI$C_\ell$ & EFTBOSS & DESI BAO & DES Y5 & Total  \\
        \hline\hline
        $\Lambda$CDM & 30510.69 & 21.77 & 32.97 & 4.87 & 2.08 & 31.90 & 653.00 & 12.77 & 1649.18 & 32919.8\\
        CPL          & 30509.83 & 22.02 & 32.83 & 4.16 & 1.92 & 32.30 & 645.86 & 8.01  & 1637.65 & 32895.2\\
        EFTofDE        & 30508.91 & 22.08 & 32.87 & 3.54 & 2.08 & 32.38 & 645.57 & 8.00  & 1637.61 & 32893.6 \\
        \hline
    \end{tabular}
    \caption{Best-fit $\chi^2$ of the various likelihoods included in our full data combination analysis (\textit{i.e.}, the ``All'' analysis) for $\Lambda$CDM, CPL with PPF, and CPL with EFTofDE.}
    \label{tab:bestfit_params_eftde_transposed}
\end{table*}

\bibliographystyle{apsrev4-1}
\bibliography{bib} 

\end{document}